\documentclass[draftcls, onecolumn]{IEEEtran}
\ifCLASSINFOpdf
\else
\fi
\usepackage{slashbox}
\usepackage{graphicx}
\usepackage{amsmath}
\usepackage{amssymb}
\usepackage{graphics}
\usepackage{latexsym}
\usepackage[dvips]{epsfig}
\usepackage[nospace]{cite}
\usepackage{subfigure}
\usepackage{setspace}
\usepackage{tabularx}
\usepackage{color}
\usepackage{arydshln}

\newtheorem{Definition}{Definition}
\newtheorem{Lemma}{Lemma}
\newtheorem{Corollary}[Lemma]{Corollary}

\newtheorem{Theorem}{Theorem}

\newtheorem{Example}{Example}
\newtheorem{Remark}{Remark}

\newcommand\blue[1]{\textcolor{blue}{#1}}
\newcommand\magenta[1]{\textcolor{magenta}{#1}}

\usepackage[colorlinks = true,
 linkcolor = blue,
 urlcolor = blue,
 citecolor = red,
 anchorcolor = green,
]{hyperref}

\allowdisplaybreaks[1]

\begin{document}
%
\title{\huge Optimal Multiplexed Erasure Codes for Streaming Messages with Different Decoding Delays}


\author{Silas L.~Fong, Ashish Khisti, Baochun Li, Wai-Tian Tan, Xiaoqing Zhu, and John Apostolopoulos%
\thanks{S.~L.~Fong is with Qualcomm Flarion Technologies, NJ 08807, USA  (E-mail: \texttt{silas.fong@ieee.org}).}
\thanks{A.~Khisti and B.~Li are with the Department of Electrical and Computer Engineering, University of Toronto, Toronto, ON M5S 3G4, Canada  (E-mails: \texttt{akhisti@ece.utoronto.ca}, \texttt{bli@ece.utoronto.edu}).}
\thanks{W.-T.~Tan, X.~Zhu and J.~Apostolopoulos are with the Enterprise Networking Innovation Labs, Cisco Systems, San Jose, CA 95134, USA.}
\thanks{This work was presented in part at 2019 IEEE International Symposium on Information Theory (ISIT).}
}


%


\maketitle

\begin{abstract}
This paper considers multiplexing two sequences of messages with two different decoding delays over a packet erasure channel. In each time slot, the source constructs a packet based on the current and previous messages and transmits the packet, which may be erased when the packet travels from the source to the destination. The destination must perfectly recover every source message in the first sequence subject to a decoding delay $T_\mathrm{v}$ and every source message in the second sequence subject to a shorter decoding delay $T_\mathrm{u}\le T_\mathrm{v}$.
We assume that the channel loss model introduces a burst erasure of a fixed length $B$ on the discrete timeline. Under this channel loss assumption, the capacity region for the case where $T_\mathrm{v}\le T_\mathrm{u}+B$ was previously solved. In this paper, we fully characterize the capacity region for the remaining case $T_\mathrm{v}> T_\mathrm{u}+B$. The key step in the achievability proof is achieving the non-trivial corner point of the capacity region through using a multiplexed streaming code constructed by superimposing two single-stream codes. The main idea in the converse proof is obtaining a genie-aided bound when the channel is subject to a periodic erasure pattern where each period consists of a length-$B$ burst erasure followed by a length-$T_\mathrm{u}$ noiseless duration.
\end{abstract}


%
\IEEEpeerreviewmaketitle

\section{Introduction} \label{Introduction}
Video streaming applications including video conferencing, virtual reality (VR) and online gaming are expected to dominate 82 percent of the Internet traffic by 2022, up from 75 percent in 2017~\cite{CiscoWhitePaper2018}. Since the user experience for a video streaming application is directly impacted by the latency and reliability guarantees supported by the underlying connection, we are motivated to find effective error correction strategies for general low-latency applications over the Internet including video streaming.

Two main error control schemes have been implemented at the data link layer and the transport layer to alleviate the effect of packet losses on applications that are run over the Internet: Automatic repeat request (ARQ) and forward error correction (FEC). In order to implement error correction for low-latency applications, FEC is preferred over ARQ when retransmitting lost packets is costly. Consider the example of remotely controlling a critical device over the Tactile Internet~\cite{tactileInternet16} where a sensor wants to communicate with an actuator in real time through a control server with round-trip latency less than 1~ms as illustrated in~\cite[Fig.~3]{5GPPP}. The latency goals for processing delay at the terminals, transmission delay over the air interfaces between the terminals and the control server and data processing delay at the control server are 0.3~ms, 0.2~ms and 0.5~ms respectively. If an ARQ scheme is used for error control, then retransmissions compete the precious time resources with data computation at the terminals and the control server. The advantage of FEC over ARQ is most obvious when retransmitting lost packets directly affects the quality of service. For example, retransmitting a Voice-over-IP (VoIP) packet incurs an extra round-trip delay (backward + forward) which will result in an overall three-way delay (forward + backward + forward) that may exceed the 150 ms delay recommended by International Telecommunication Union~\cite{onewayTransTime} (see~\cite{StockhammerHannuksela2005} for an overview of the ubiquitous H.264/AVC video coding standard). Given the fact that the three-way propagation delay (forward + backward + forward) is at least 200~ms for communication between two diametrically opposite points on the earth's circumference~\cite{BKTAmagazine17}, FEC has a clear advantage over ARQ for long-distance low-latency communication.

This paper focuses on low-latency FEC schemes implemented at the transport layer, where a source packet is either received by the destination without error or dropped by the network (possibly due to unreliable links or network congestion). In other words, a source packet is either perfectly recovered by the destination or completely erased. Since packet erasures often occur in a bursty rather than sparse manner~\cite{Bolot1993,Paxson1999}, we model the connection between the source and the destination as a packet erasure channel that introduces burst erasures. In order to capture the nature of streaming messages and the low-latency requirements, we assume that a source message is generated in every time slot and a decoding delay constraint~$T$ is imposed on every message, where each message is encoded into a channel packet before being transmitted through the erasure channel. If the destination cannot decode a message within~$T$ time slots from the time when the message is generated, the message is considered lost. Ideally, we would like to characterize the maximum achievable rates for statistical models that generate burst erasures such as the well-known Gilbert-Elliott channel \cite{Gilbert1960,Elliott1963} and its generalization the Fritchman channel~\cite{Fritchman1967}. However, such characterizations seem intractable due to the decoding delay constraint and the fact that statistical models that generate burst erasures are not memoryless. Therefore, Martinian and Sundberg~\cite{MartinianSundberg2004} have instead fully characterized the capacity, i.e., maximum coding rate, for a simpler deterministic model where a burst erasure of length~$B$ is introduced on the discrete timeline and every message has to be perfectly recovered at the destination with a decoding delay of~$T$ time slots. They proposed a streaming code that not only achieves the capacity~$\frac{T}{T+B}$ for the deterministic model, but also can significantly outperform traditional FEC schemes for the Gilbert-Elliott channel. Various generalizations of the packet erasure model and the streaming codes studied in~\cite{MartinianSundberg2004} have been proposed in~\cite{LeungHo2012, LQH2013, BKTA2013, AdlerCassuto2017,FKLTZA2018}.

Note that an Internet application may consist of multiple types of data streams (video, audio, text, etc.), and also within a single data stream such as video there are different subsets of data that have different delivery deadlines. Moreover, multiplexing streams of different latency constraints has been implemented in the QUIC transport protocol to reduce latency of Google Search responses and reduce rebuffer rates of YouTube playbacks~\cite{QUIC2017}. Therefore, Badr et al.\ \cite{BLKTZA2018} extended the study of single-stream codes in~\cite{MartinianSundberg2004} and initiated the study of streaming codes which multiplex a stream of urgent messages with a stringent delay constraint and a stream of less-urgent messages with a less stringent delay constraint. Simulation results in~\cite[Sec.~VIII]{BLKTZA2018} demonstrate that using multiplexed streaming codes can significantly outperform concatenating multiple single-stream codes for the Gilbert-Elliott channel. In the multiplexed streaming model studied in~\cite{BLKTZA2018}, every urgent message has to be decoded within $T_\mathrm{u}$ time slots from the time when the urgent message is generated, and every less-urgent message has to be decoded within $T_\mathrm{v}$ time slots from the time when the less-urgent message is generated. It is assumed that $T_\mathrm{u}\le T_\mathrm{v}$, consistent with the notion that the urgent messages have to be decoded with less decoding delay than the less-urgent messages. Similar to the single-stream case, we assume that the channel introduces a burst erasure of length~$B$ on the discrete timeline and define the capacity region to be the set of rate pairs $(R_\mathrm{v}, R_\mathrm{u})$ which are supported by streaming codes that correct any length-$B$ burst erasure where $R_\mathrm{v}$ and $R_\mathrm{u}$ denote the rates of the less-urgent stream and urgent stream respectively. For the case $T_\mathrm{u}\le T_\mathrm{v}\le T_\mathrm{u}+B$, systematic streaming codes have been proposed in~\cite{BLKTZA2018} to achieve the capacity region.
However, for the remaining non-trivial case $T_\mathrm{v}> T_\mathrm{u}+B$, it is unclear whether the capacity region can be achieved by the multiplexed streaming codes proposed in~\cite{BLKTZA2018}. Therefore, we are motivated to investigate the capacity region for the case $T_\mathrm{v}> T_\mathrm{u}+B$.

\subsection{System Model}\label{subsecChannelModel}
In order to describe the existing results for the packet-erasure channel model, we would like to briefly describe the channel model. A formal description will appear later in the paper. The channel consists of a source and a destination. In each time slot, the source generates a collection of~$k_\mathrm{u}$ urgent symbols and a collection of~$k_\mathrm{v}$ less-urgent symbols. Then, the source encodes the $k_\mathrm{u}+k_\mathrm{v}$ symbols into a collection of~$n$ symbols followed by transmitting the~$n$ symbols through the channel. The collection of~$n$ symbols transmitted in a time slot are either received perfectly by the destination or erased (lost). The fractions $k_{\textrm{u}}/n$ and $k_{\mathrm{v}}/n$ specify the rates of the urgent and less-urgent streams respectively. We call the~$k_\mathrm{u}$ symbols chosen by the source, the~$k_\mathrm{v}$ symbols chosen by the source, the~$n$ symbols transmitted by the source and the~$n$ symbols received by the destination the \emph{urgent source packet}, the \emph{less-urgent source packet}, the \emph{transmitted packet} and the \emph{received packet} respectively. We assume that every urgent source packet generated in a time slot must be decoded with delay~$T_\mathrm{u}$, i.e., within the future~$T_\mathrm{u}$ time slots, and every less-urgent source packet generated in a time slot must be decoded with delay
\begin{equation}
T_\mathrm{v}\ge T_\mathrm{u}. \label{TvGreaterThanTu}
\end{equation}

In order to capture the packet loss behavior over the Internet, we consider the simple scenario where the channel introduces on the discrete timeline a burst erasure of length~$B$.
We assume without loss of generality (wlog) that
\begin{align}
T_\mathrm{v}\ge B, \label{TgreaterThanB}
\end{align}
or otherwise a burst erasure of length~$B$ starting from a certain time slot would prevent the destination to timely recover (within $T_\mathrm{v}$ time slots) both the urgent and less-urgent source packets generated in the same time slot. If the channel is noiseless where $B=0$, no coding is needed to asymptotically achieve all the rate pairs $(k_{\mathrm{v}}/n, k_{\textrm{u}}/n)$ on the boundary of the capacity region that satisfy $k_{\textrm{u}}/n + k_{\mathrm{v}}/n = 1$. Therefore, we assume wlog that
\begin{align}
B\ge 1. \label{assumptionIntro}
\end{align}
\subsection{Existing Results} \label{subsecKnownResults}
For the case $T_\mathrm{u}< B$ under the assumption $T_\mathrm{v}\ge B\ge 1$ by~\eqref{TgreaterThanB} and~\eqref{assumptionIntro}, it can be observed that a burst erasure of length~$B$ starting from a certain time slot would prevent the destination to timely recover (within $T_\mathrm{u}$ time slots) the urgent source packet transmitted in the same time slot. Consequently, no rate pair $(k_{\mathrm{v}}/n, k_{\textrm{u}}/n)$ with $k_{\textrm{u}}/n>0$ is achievable, which implies that the capacity region reduces to the interval $[0, \mathrm{C}(T_\mathrm{v}, B)]$ on the horizontal axis where
\begin{align}
\mathrm{C}(T, B) \triangleq \frac{T}{T+B} \label{defCapacityPTP}
\end{align}
denotes the maximum coding rate of streaming codes with delay~$T$ that correct any length-$B$ burst erasure~\cite[Th.~1 and Th.~2]{MartinianSundberg2004} (see also~\cite[Sec.~III-C]{FKLTZA2018}). Since the case $T_\mathrm{u}< B$ degenerates the multiplexing problem to the previously known single-stream problem described above, we assume wlog that
\begin{align}
T_\mathrm{u} \ge B. \label{TuGreaterThanB}
\end{align}

For the case $T_\mathrm{u}=T_\mathrm{v}$, since the urgent and less-urgent source packets can be viewed as single-stream source packets with delay $T_\mathrm{u}$, any rate pair $(k_{\mathrm{v}}/n, k_{\textrm{u}}/n)$ must satisfy $k_{\textrm{u}}/n + k_{\mathrm{v}}/n \le \mathrm{C}(T_\mathrm{u}, B)$ (recall that the capacity of the single-stream problem equals $\mathrm{C}(T, B)$ by~\cite{MartinianSundberg2004}). In addition, the boundary of the capacity region $k_{\textrm{u}}/n + k_{\mathrm{v}}/n = \mathrm{C}(T_\mathrm{u}, B)$ can be asymptotically achieved by partitioning each source packet of an optimal code with rate $\mathrm{C}(T_\mathrm{u}, B)$ into an urgent source packet and a less-urgent source packet. Consequently, the case $T_\mathrm{u}= T_\mathrm{v}$ degenerates the multiplexing problem to a single-stream problem described above. Therefore, in view of~\eqref{TvGreaterThanTu}, we assume wlog that
\begin{align}
T_\mathrm{v}>T_\mathrm{u}.  \label{TuSmallerThanTv}
\end{align}
Summarizing the assumptions \eqref{assumptionIntro}, \eqref{TuGreaterThanB} and~\eqref{TuSmallerThanTv}, we assume in the rest of the paper that
\begin{align}
T_\mathrm{v}> T_\mathrm{u} \ge B \ge 1. \label{assumptionWholePaper}
\end{align}
Any condition that does not satisfy~\eqref{assumptionWholePaper} leads to known results as explained in this and the previous subsections.

\begin{figure}[!t]
\centering
   \subfigure[Case $T_\mathrm{u}< T_\mathrm{v}\le T_\mathrm{u}+B$]{
        \includegraphics[width=3 in]{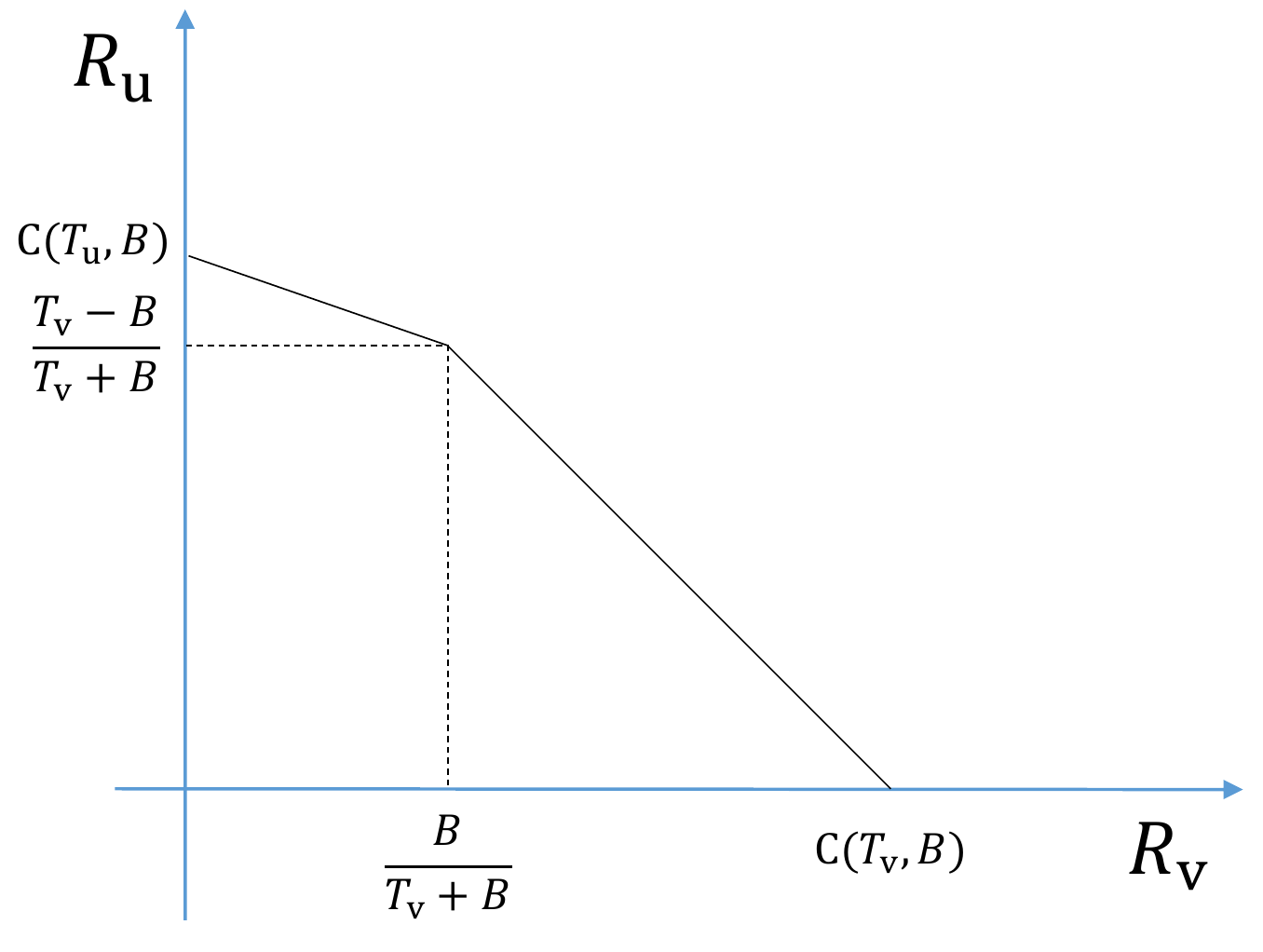}

   }
\subfigure[Case $T_\mathrm{v}> T_\mathrm{u}+B$]{
        \includegraphics[width=3 in]{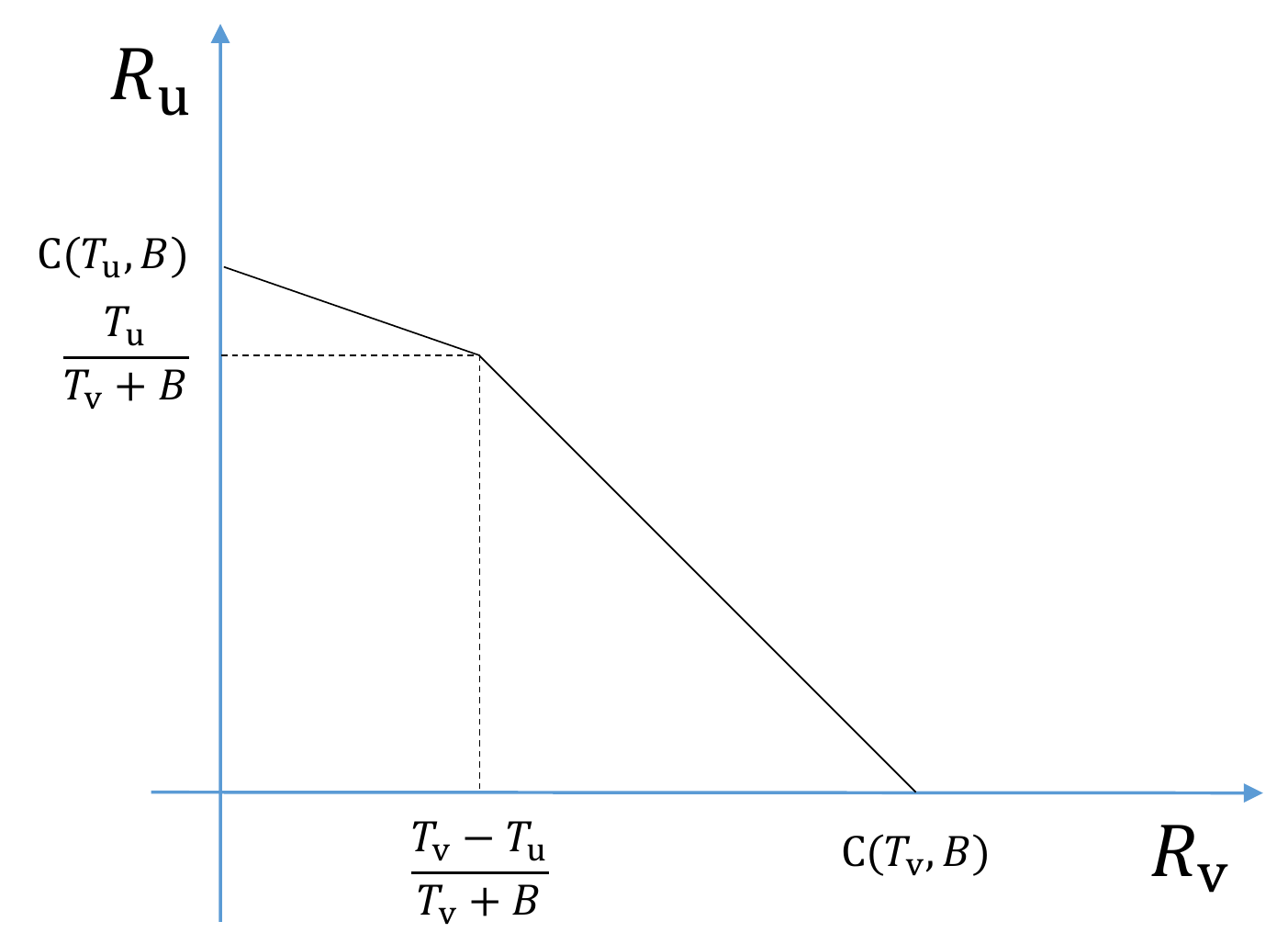}

   }
\caption{Capacity region}
\label{figureCapacity}
\end{figure}

For the special case where
\begin{align}
1\le B\le T_\mathrm{u}< T_\mathrm{v} \le T_\mathrm{u}+B, \label{case1Capacity}
\end{align}
  systematic streaming codes have been proposed in~\cite{BLKTZA2018} to achieve the capacity region, which is the set of rate pairs $(R_{\mathrm{v}}, R_{\mathrm{u}})$ satisfying
\begin{align*}
\left(1+\frac{T_\mathrm{u}+B-T_\mathrm{v}}{T_\mathrm{u}}\right)R_{\mathrm{v}} + \frac{R_{\mathrm{u}}}{\mathrm{C}(T_\mathrm{u}, B)} \le 1
\end{align*}
and
\begin{align}
R_{\mathrm{v}} + R_{\mathrm{u}} \le \mathrm{C}(T_\mathrm{v}, B) \label{capacityEq2}
\end{align}
as illustrated in Figure~\ref{figureCapacity}(a). In addition, other systematic streaming codes have been proposed in~\cite{BLKTZA2018} to achieve two different rate regions for the cases $T_\mathrm{u}+B< T_\mathrm{v}<T_\mathrm{u}+2B $ and $T_\mathrm{v}\ge  T_\mathrm{u}+2B$ respectively, denoted by $\mathcal{R}_{\{T_\mathrm{u}+B< T_\mathrm{v}<T_\mathrm{u}+2B\}}$ and $\mathcal{R}_{\{T_\mathrm{v}\ge  T_\mathrm{u}+2B\}}$ respectively. In particular, if only systematic streaming codes are allowed, $\mathcal{R}_{\{T_\mathrm{v}\ge  T_\mathrm{u}+2B\}}$ was shown in~\cite{BLKTZA2018} to be the largest.

\subsection{Main Contribution}
  Under the assumption~\eqref{assumptionWholePaper}, the capacity region for case~\eqref{case1Capacity} was proved in~\cite{BLKTZA2018}. This paper solves the only remaining case
\begin{align}
 T_\mathrm{v}> T_\mathrm{u}+B \label{assumptionWholePaper*}
\end{align}
and characterize the capacity region to be the set of rate pairs $(R_{\mathrm{v}}, R_{\mathrm{u}})$ satisfying~\eqref{capacityEq2}
and
\begin{align}
R_{\mathrm{v}} + \frac{R_{\mathrm{u}}}{\mathrm{C}(T_\mathrm{u}, B)} \le 1  \label{capacityEq3}
\end{align}
 as illustrated in Figure~\ref{figureCapacity}(b).

In order to prove the achievability, we propose a \emph{non-systematic} streaming code that achieves the non-trivial corner point $\big(\frac{T_\mathrm{v}-T_\mathrm{u}}{T_\mathrm{v}+B},\frac{T_\mathrm{u}}{T_\mathrm{v}+B}\big)$. The proposed multiplexed streaming code is constructed by superimposing two single-stream codes with respective rates~$\frac{T_\mathrm{v}-T_\mathrm{u}}{T_\mathrm{v}-T_\mathrm{u}+B}$ and $\frac{T_\mathrm{u}}{T_\mathrm{u}+B}$ and respective delays~$T_\mathrm{v}-T_\mathrm{u}$ and~$T_\mathrm{u}$.
In order to prove the converse, we first prove a genie-aided outer bound when the channel is subject to a periodic erasure pattern where each period consists of a length-$B$ burst erasure followed by a length-$T_\mathrm{u}$ noiseless duration. The genie provides the least amount of information to the destination so that both the urgent and less-urgent streams can be perfectly recovered at the destination. {Then, we average the genie-aided bound over all offsets of the periodic erasure pattern and combine the averaged genie-aided bound with the existing trivial bound $R_{\mathrm{v}} + R_{\mathrm{u}}\le \mathrm{C}(T_\mathrm{v}, B)$,} resulting an outer bound with four corner points as shown in Figure~\ref{figureCapacity}(b). In particular, for the case $T_\mathrm{v}\ge  T_\mathrm{u}+2B$, the converse proof combined with the result in~\cite{BLKTZA2018} as described at the end of Section~\ref{subsecKnownResults} implies that systematic streaming codes alone are sufficient to achieve the capacity region.

\subsection{Paper Outline}
This paper is organized as follows. The notation in this paper is explained in the next subsection. Section~\ref{sectionDefinition} presents the formulation of multiplexed streaming codes for the packet erasure channel and states the main result --- the capacity region for the case $T_\mathrm{v}>T_\mathrm{u}+B$. Section~\ref{sectionAchievability} contains the achievability proof of the main result which involves the construction of a multiplexed streaming code that achieves the non-trivial corner point of the capacity region. Section~\ref{sectionConverse} presents the converse proof of the main result which involves obtaining a genie-aided bound.
Section~\ref{conclusion} concludes this paper.
\subsection{Notation}\label{notation}
The sets of natural numbers, integers, non-negative integers, and non-negative real numbers are denoted by $\mathbb{N}$, $\mathbb{Z}$, $\mathbb{Z}_+$ and $\mathbb{R}_+$ respectively. All the elements of any matrix considered in this paper are taken from a common finite field~$\mathbb{F}$, where~$0$ and~$1$ denote the additive identity and the multiplicative identity respectively. The set of $k$-dimensional row vectors over $\mathbb{F}$ is denoted by $\mathbb{F}^k$, and the set of $k\times n$ matrices over $\mathbb{F}$ is denoted by $\mathbb{F}^{k\times n}$. A row vector in $\mathbb{F}^k$ is denoted by $\mathbf{a} \triangleq [a_0\ a_1\ \ldots \ a_{k-1}]$ where $a_\ell$ denotes the $(\ell+1)^{\text{th}}$ element of $\mathbf{a}$.
The $k$-dimensional identity matrix is denoted by $\mathbf{I}_k$
and the $L\times B$ all-zero matrix is denoted by $\mathbf{0}^{L\times B}$. An $L \times B$ parity matrix of a systematic maximum-distance separable (MDS) $(L+B, L)$-code is denoted by
$\mathbf{V}^{L\times B}$,
which possesses the property that any $L$ columns of $[\mathbf{I}_{L} \ \mathbf{V}^{L\times B}]\in \mathbb{F}^{L\times (L+B)}$ are independent. It is well known that a systematic maximum-distance separable (MDS) $(L+B, L)$-code always exists as long as $|\mathbb{F}|\ge L+B$ \cite{MacWilliamsSloane1988}. We will take all logarithms to base~$2$ throughout this paper.  {For any discrete random tuple $(X,Y,Z)$, we let $H(X|Z)$ denote the entropy of $X$ given $Z$, and let $I(X;Y|Z)$ denote the mutual information between $X$ and $Y$ given~$Z$.}

\section{Multiplexed Streaming Codes for Channels with Burst Erasures} \label{sectionDefinition}
\subsection{Problem formulation}
The source wants to simultaneously send a sequence of length-$k_{\textrm{u}}$ packets~$\mathbf{u}^\infty\triangleq \{\mathbf{u}_i\}_{i=0}^{\infty}$ with decoding delay $T_\mathrm{u}$ and a sequence of length-$k_{\mathrm{v}}$ packets~$\mathbf{v}^\infty\triangleq\{\mathbf{v}_i\}_{i=0}^{\infty}$ with decoding delay $T_\mathrm{v}\ge T_\mathrm{u}$ to the destination, where $k_{\textrm{u}}$ and  $k_{\mathrm{v}}$ denote the sizes of each urgent packet $\mathbf{u}_i$ and each less-urgent packet $\mathbf{v}_i$ respectively. Each $\mathbf{u}_i$ is an element in $\mathbb{F}^{k_{\textrm{u}}}$ and each $\mathbf{v}_i$ is an element in $\mathbb{F}^{k_{\mathrm{v}}}$ where $\mathbb{F}$ is some finite field. In each time slot~$i\in\mathbb{Z}_+$, the source packets $\mathbf{v}_i$ and $\mathbf{u}_i$ are encoded into a length-$n$ packet $\mathbf{x}_i\in\mathbb{F}^n$ to be transmitted to the destination through an erasure channel, and the destination receives $\mathbf{y}_i\in\mathbb{F}^n \cup \{*\}$ where the received packet $\mathbf{y}_i$ equals either the transmitted packet $\mathbf{x}_i$ or the erasure symbol~`$*$'. The urgent and less-urgent streams are subject to the delay constraints of~$T_\mathrm{u}$ and~$T_\mathrm{v}$ time slots respectively, meaning that the destination must produce an estimate of $\mathbf{u}_i$, denoted by $\hat{\mathbf{u}}_i$, upon receiving $\mathbf{y}_{i+T_\mathrm{u}}$ and produce an estimate of $\mathbf{v}_i$, denoted by $\hat{\mathbf{v}}_i$, upon receiving $\mathbf{y}_{i+T_\mathrm{v}}$.
As mentioned in Section~\ref{subsecChannelModel}, we assume that the channel introduces a burst erasure of length~$B$. Recall that we assume~\eqref{assumptionWholePaper} wlog.

%
\subsection{Standard definitions}
\begin{Definition} \label{definitionCode}
An $(n, k_\mathrm{v}, k_\mathrm{u}, T_\mathrm{v}, T_\mathrm{u})_{\mathbb{F}}$-streaming code consists of the following:
\begin{enumerate}
\item A sequence of less-urgent source packets~$\mathbf{v}^{\infty}$ where $\mathbf{v}_i\in \mathbb{F}^{k_\mathrm{v}}$.
\item A sequence of urgent source packets~$\mathbf{u}^{\infty}$ where $\mathbf{u}_i\in \mathbb{F}^{k_\mathrm{u}}$.
\item 
    An encoder~$f_i: \underbrace{\mathbb{F}^{k_\mathrm{u} + k_\mathrm{v}} \times \ldots \times \mathbb{F}^{k_\mathrm{u} + k_\mathrm{v}}}_{i +1 \text{ times }} \rightarrow \mathbb{F}^n$ for each $i\in\mathbb{Z}_+$, where $f_i$ is used by the source at time~$i$ to encode $\mathbf{u}_i$ and $\mathbf{v}_i$ such that
    \[
    \mathbf{x}_i = f_i((\mathbf{u}_0, \mathbf{v}_0), (\mathbf{u}_1, \mathbf{v}_1), \ldots, (\mathbf{u}_i, \mathbf{v}_i)).
    \]

         \item A decoding function~$\varphi_{i+T_\mathrm{v}}^{(\mathrm{v})}: \underbrace{\mathbb{F}^n\cup\{*\} \times \ldots \times \mathbb{F}^n\cup\{*\}}_{i+T_\mathrm{v}+1 \text{ times }}\rightarrow \mathbb{F}^{k_\mathrm{v}}$ for each $i\in\mathbb{Z}_+$, where $\varphi_{i+T_\mathrm{v}}^{(\mathrm{v})}$ is used by the destination at time $i+T_\mathrm{v}$ to estimate $\mathbf{v}_i$ such that
        \begin{align}
      \hat{\mathbf{v}}_i= \varphi_{i+T_\mathrm{v}}^{(\mathrm{v})}(\mathbf{y}_0, \mathbf{y}_1, \ldots, \mathbf{y}_{i+T_\mathrm{v}}). \label{defDecoderNonurgent}
        \end{align}
         \item A decoding function~$\varphi_{i+T_\mathrm{u}}^{(\mathrm{u})}: \underbrace{\mathbb{F}^n\cup\{*\} \times \ldots \times \mathbb{F}^n\cup\{*\}}_{i+T_\mathrm{u}+1 \text{ times }}\rightarrow \mathbb{F}^{k_\mathrm{u}}$ for each $i\in\mathbb{Z}_+$, where $\varphi_{i+T_\mathrm{u}}^{(\mathrm{u})}$ is used by the destination at time $i+T_\mathrm{u}$ to estimate $\mathbf{u}_i$ according to
        \begin{align}
      \hat{\mathbf{u}}_i= \varphi_{i+T_\mathrm{u}}^{(\mathrm{u})}(\mathbf{y}_0, \mathbf{y}_1, \ldots, \mathbf{y}_{i+T_\mathrm{u}}). \label{defDecoderUrgent}
        \end{align}
\end{enumerate}
In addition, the code is said to be $\emph{systematic}$ if $\mathbf{x}_i=[\mathbf{v}_i\ \mathbf{u}_i\ \mathbf{a}_i]$ for some $\mathbf{a}_i\in\mathbb{F}^{n-k_\mathrm{v}-k_\mathrm{u}}$ at each time $i\in\mathbb{Z}_+$.
\end{Definition}

The formal definition of a length-$B$ burst erasure is given below.
\begin{Definition} \label{definitionErasureSeq}
An erasure sequence is a binary sequence denoted by $\mathbf{e}\triangleq\{e_i\}_{i=0}^{\infty}$ where
\begin{equation*}
e_i=\mathbf{1}\{\text{erasure occurs at time~$i$}\}.
\end{equation*}
If $\sum_{i=0}^\infty e_i= B$ holds with all the~$1$'s occupying consecutive positions, $\mathbf{e}$ is called a $B$-erasure sequence. The set of $B$-erasure sequences is denoted by $\Omega_B$. Similarly, for any~$n\ge B$, a length-$n$ binary sequence denoted by $e^n\triangleq\{e_i\}_{i=0}^{n-1}$ is called a $B$-erasure sequence if $e^n$ satisfies $\sum_{i=0}^{n-1} e_i= B$ with all the~$1$'s occupying consecutive positions. The set of length-$n$ $B$-erasure sequences is denoted by $\Omega_B^n$.
\end{Definition}

\medskip
\begin{Definition} \label{definitionChannel}
The mapping $g_n: \mathbb{F}^n \times \{0,1\} \rightarrow \mathbb{F}^n \cup \{*\}$ of the erasure channel is defined as
\begin{align}
g_n(\mathbf{x},e)= \begin{cases}\mathbf{x} & \text{if $e=0$,}\\
* & \text{if $e=1$.}
\end{cases} \label{defChannelLaw}
\end{align}
For any erasure sequence $\mathbf{e}$ and any $(n, k_\mathrm{v}, k_\mathrm{u}, T_\mathrm{v}, T_\mathrm{u})_{\mathbb{F}}$-streaming code, the following input-output relation holds for the erasure channel for each $i\in\mathbb{Z}_+$:
\begin{align}
\mathbf{y}_i = g_n(\mathbf{x}_i, e_i).\label{defChannelOutput}
\end{align}
\end{Definition}
\medskip
\begin{Definition} \label{definitionRecover}
An $(n, k_\mathrm{v}, k_\mathrm{u}, T_\mathrm{v}, T_\mathrm{u})_{\mathbb{F}}$-streaming code is said to correct a $B$-erasure sequence $\mathbf{e}\in\Omega_B$ if the following holds: For all $i\in\mathbb{Z}_+$ and all $[\mathbf{u}_i\ \mathbf{v}_i]\in \mathbb{F}^{k_\mathrm{u}+k_\mathrm{v}}$, we have
\begin{equation*}
 [\hat{\mathbf{u}}_i\ \hat{\mathbf{v}}_i] = [\mathbf{u}_i\ \mathbf{v}_i]
\end{equation*}
where
\begin{equation*}
  \hat{\mathbf{u}}_i = \varphi_{i+T_\mathrm{u}}^{(\mathrm{u})}\big( g_n(\mathbf{x}_0, e_0), \ldots,  g_n(\mathbf{x}_{i+T_\mathrm{u}}, e_{i+T_\mathrm{u}})\big)
\end{equation*}
and
\begin{equation*}
\hat{\mathbf{v}}_i = \varphi_{i+T_\mathrm{v}}^{(\mathrm{v})}\big( g_n(\mathbf{x}_0, e_0), \ldots,  g_n(\mathbf{x}_{i+T_\mathrm{v}}, e_{i+T_\mathrm{v}})\big)
\end{equation*}
due to~\eqref{defDecoderUrgent}, \eqref{defDecoderNonurgent} and~\eqref{defChannelOutput}.
\end{Definition}
\medskip

{The following corollary is a direct consequence of Definition~\ref{definitionCode} and Definition~\ref{definitionRecover}, and its proof is relegated to Appendix~\ref{appendixA-}.
\begin{Corollary}\label{corollaryConcatenate}
Suppose an $(n, k_\mathrm{v}, k_\mathrm{u}, T_\mathrm{v}, T_\mathrm{u})_{\mathbb{F}}$-streaming code that corrects any $B$-erasure sequence exists. Then for each $q\in\mathbb{N}$, we can construct a $(qn, qk_\mathrm{v}, qk_\mathrm{u}, T_\mathrm{v}, T_\mathrm{u})_{\mathbb{F}}$-streaming code that corrects any $B$-erasure sequence.
\end{Corollary}
}
\medskip
\begin{Definition} \label{definitionAchievability}
A rate pair $(R_\mathrm{v}, R_\mathrm{u})\in\mathbb{R}_+^2$ is said to be $(T_\mathrm{v}, T_\mathrm{u},B)$-achievable if there exists an $(n, k_\mathrm{v}, k_\mathrm{u}, T_\mathrm{v}, T_\mathrm{u})_{\mathbb{F}}$-streaming code which corrects any $B$-erasure sequence such that
$
 \frac{k_\mathrm{v}}{n} \ge R_\mathrm{v} 
$
and
$
\frac{k_\mathrm{u}}{n} \ge R_\mathrm{u} 
$.
\end{Definition}
\medskip

The following corollary is a direct consequence of Definition~\ref{definitionAchievability} and the existing single-stream result~\cite[Th.~2]{MartinianSundberg2004} (see also~\cite[Th.~1]{FKLTZA2018})) stated as follows: Suppose $T\ge B\ge 1$. Then, there exists a streaming code with rate~$\mathrm{C}(T,B)$ which guarantees the recovery of every streaming message with delay~$T$ when the channel is subject to any length-$B$ burst erasure on the discrete timeline. 
\begin{Corollary}[{\cite[Th.~2]{MartinianSundberg2004}}] \label{corollaryAch}
The rate pairs $(\mathrm{C}(T_\mathrm{v},B),0)$ and $(0, \mathrm{C}(T_\mathrm{u},B))$ are  $(T_\mathrm{v}, T_\mathrm{u},B)$-achievable.
\end{Corollary}
\medskip
\begin{Definition} \label{definitionCapacity}
Fix any $(T_\mathrm{v}, T_\mathrm{u},B)$ that satisfies~\eqref{assumptionWholePaper}. The $(T_\mathrm{v}, T_\mathrm{u},B)$-achievable rate region, denoted by $\mathcal{C}_{T_\mathrm{v}, T_\mathrm{u},B}$, is the closure of the set of $(T_\mathrm{v}, T_\mathrm{u},B)$-achievable rate pairs.
\end{Definition}
\medskip

The following convexity statement regarding~$\mathcal{C}_{T_\mathrm{v}, T_\mathrm{u},B}$ will help us simplify the achievability proof of our main result. The proof is standard and is therefore relegated to Appendix~\ref{appendixA}.
\begin{Corollary} \label{corollaryConvex}
For any $(T_\mathrm{v}, T_\mathrm{u},B)$ that satisfies~\eqref{assumptionWholePaper}, $\mathcal{C}_{T_\mathrm{v}, T_\mathrm{u},B}$ is convex.
\end{Corollary}

\subsection{Main Result} \label{subsecMainResult}
The following theorem is the main result of this paper, which states the capacity region in terms of the single-stream capacity function~$\mathrm{C}(\cdot, \cdot)$ as defined in~\eqref{defCapacityPTP}.
\begin{Theorem}\label{thmMainResult}
Fix any $(T_\mathrm{v},T_\mathrm{u},B)$ that satisfies~\eqref{assumptionWholePaper} and~\eqref{assumptionWholePaper*}. Define
\begin{align}
\mathcal{R}_{\{T_\mathrm{v}> T_\mathrm{u}+B\}}\triangleq \left\{(R_\mathrm{v}, R_\mathrm{u})\in\mathbb{R}_+^2 \left|\: \parbox[c]{1.2 in}{$
R_{\mathrm{v}} + \frac{R_{\mathrm{u}}}{\mathrm{C}(T_\mathrm{u}, B)} \le 1,
\\
R_{\mathrm{v}} + R_{\mathrm{u}} \le \mathrm{C}(T_\mathrm{v}, B)$}\right.  \right\} \label{defSetR}
\end{align}
as illustrated in Figure~\ref{figureCapacity}(b).
Then,
\begin{align*}
\mathcal{C}_{T_\mathrm{v}, T_\mathrm{u},B} =
\mathcal{R}_{\{T_\mathrm{v}> T_\mathrm{u}+B\}}.
\end{align*}
\end{Theorem}
\begin{Remark} \label{remark1}
Consider the special case where $T_\mathrm{v} \ge T_\mathrm{u} +2B$. It has been shown in~\cite[Th.~1]{BLKTZA2018} that systematic streaming codes (cf.\ Definition~\ref{definitionCode}) achieve~$\mathcal{R}_{\{T_\mathrm{v}> T_\mathrm{u}+B\}}$. Therefore, it follows from Theorem~\ref{thmMainResult} that systematic streaming codes are sufficient to achieve the capacity region.
\end{Remark}
\begin{Remark} \label{remark2}
Consider the special case where $T_\mathrm{u}+B< T_\mathrm{v} < T_\mathrm{u} +2B$. The systematic streaming codes proposed in~\cite[Th.~1]{BLKTZA2018} cannot achieve the non-trivial corner point $\big(\frac{T_\mathrm{v}-T_\mathrm{u}}{T_\mathrm{v}+B},\frac{T_\mathrm{u}}{T_\mathrm{v}+B}\big)$ of the capacity region $\mathcal{C}_{T_\mathrm{v}, T_\mathrm{u},B}$. On the other hand, our achievability proof presented in Section~\ref{sectionAchievability} proposes a \emph{non-systematic} streaming code that achieves the non-trivial corner point. It remains open whether systematic streaming codes are sufficient to achieve the capacity region.
\end{Remark}

\section{Achievability Proof of Main Result} \label{sectionAchievability}
The achievability proof of Theorem~\ref{thmMainResult} consists of two steps. The first step involves constructing a multiplexed block code which corrects any $B$-erasure sequence. The second step involves constructing a multiplexed streaming code which corrects any $B$-erasure sequence by periodically interleaving the multiplexed block code. The formal definitions and existing results related to multiplexed block codes and periodic interleaving are presented in the following subsection.
\subsection{Preliminaries}
\begin{Definition} \label{definitionCodeBlock}
An $(n, k_\mathrm{v}, k_\mathrm{u}, T_\mathrm{v}, T_\mathrm{u})_{\mathbb{F}}$-block code consists of the following:
\begin{enumerate}
\item A vector of $k_\mathrm{v}$ less-urgent source symbols in~$\mathbb{F}$ denoted by $\vec v\triangleq\big[v[0]\ v[1]\ \ldots\ v[k_\mathrm{v}-1]\big]$.
    \item A vector of $k_\mathrm{u}$ urgent source symbols in~$\mathbb{F}$ denoted by $\vec u \triangleq \big[u[0]\ u[1]\ \ldots\ u[k_\mathrm{u}-1]\big]$.
\item
    A generator matrix $\mathbf{G}\in \mathbb{F}^{(k_\mathrm{v}+k_\mathrm{u})\times n}$. The codeword is generated according to
\begin{align*}
\big[x[0]\ x[1]\ \ldots \ x[n-1]]=[\vec v\ \vec u\big]\, \mathbf{G}. 
\end{align*}
\item A decoding function~$\varphi_{i+T_\mathrm{v}}^{(\mathrm{v})}: \underbrace{\mathbb{F}\cup\{*\} \times \ldots \times \mathbb{F}\cup\{*\}}_{i+T_\mathrm{v}+1 \text{ times }}\rightarrow \mathbb{F}$ for each $i\in\{0, 1, \ldots, k_\mathrm{v}-1\}$, where $\varphi_{i+T_\mathrm{v}}^{(\mathrm{v})}$ is used by the destination at time $i+T_\mathrm{v}$ to estimate $v[i]$ according to
        \begin{align*}
      \hat{v}[i]=
      \begin{cases}
      \varphi_{i+T_\mathrm{v}}^{(\mathrm{v})}(y[0], y[1], \ldots, y[i+T_\mathrm{v}]) & \text{if $i+T_\mathrm{v}\le n-1$}  \\  \varphi_{i+T_\mathrm{v}}^{(\mathrm{v})}(\underbrace{y[0], \ldots, y[n-1], *, \ldots, *}_{i+T_\mathrm{v}+1\text{ symbols}}) & \text{if $i+T_\mathrm{v}> n-1$.}
     \end{cases}
       \end{align*}
    \item A decoding function~$\varphi_{i+T_\mathrm{u}}^{(\mathrm{u})}: \underbrace{\mathbb{F}\cup\{*\} \times \ldots \times \mathbb{F}\cup\{*\}}_{i+T_\mathrm{u}+1 \text{ times }}\rightarrow \mathbb{F}$ for each $i\in\{0, 1, \ldots, k_\mathrm{u}-1\}$, where $\varphi_{i+T_\mathrm{u}}^{(\mathrm{u})}$ is used by the destination at time $i+T_\mathrm{u}$ to estimate $u[i]$ according to
        \begin{align*}
      \hat{u}[i]=
      \begin{cases}
      \varphi_{i+T_\mathrm{u}}^{(\mathrm{u})}(y[0], y[1], \ldots, y[i+T_\mathrm{u}]) & \text{if $i+T_\mathrm{u}\le n-1$}  \\  \varphi_{i+T_\mathrm{u}}^{(\mathrm{u})}(\underbrace{y[0], \ldots, y[n-1], *, \ldots, *}_{i+T_\mathrm{u}+1\text{ symbols}}) & \text{if $i+T_\mathrm{u}> n-1$.}
     \end{cases}
       \end{align*}
\end{enumerate}
\end{Definition}

The following definition concerns the error-correcting capability of a block code.
\begin{Definition} \label{definitionAchievabilityBlock}
 An $(n, k_\mathrm{v}, k_\mathrm{u}, T_\mathrm{v}, T_\mathrm{u})_{\mathbb{F}}$-block code is said to correct a $B$-erasure sequence $e^n\in\Omega_B^n$ if the following holds: Let $y[i]=g_1(x[i], e_i)$ be the symbol received by the destination at time~$i$ for each $i\in\{0, 1,\ldots, n-1\}$ where $g_1$ is defined in~\eqref{defChannelLaw}. Then, $\hat v[i] = v[i]$ holds for all $i\in\{0, 1, \ldots, k_\mathrm{v}-1\}$ and all $v[i]\in \mathbb{F}$, and $\hat u[i] = u[i]$ holds for all $i\in\{0, 1, \ldots, k_\mathrm{u}-1\}$ and all $u[i]\in \mathbb{F}$,
 {where
$\hat{v}[i]$
        and
        $ \hat{u}[i]$ are as defined in Definition~\ref{definitionCodeBlock}.}
\end{Definition}%

The following lemma implies that constructing a streaming code which corrects any length-$B$ burst erasure is not more difficult than constructing a block code which corrects any length-$B$ burst erasure. The proof of the following lemma is deferred to Appendix~\ref{appendixB} because it follows the standard argument of interleaving a block code into a streaming code by means of periodic interleaving~\cite{Forney1971} (see also \cite[Sec.\ IV-A]{MartinianSundberg2004}).
\begin{Lemma}\label{lemmaBlockToStreaming}
Given an $(n, k_\mathrm{v}, k_\mathrm{u}, T_\mathrm{v}, T_\mathrm{u})_{\mathbb{F}}$-block code which corrects any $B$-erasure sequence, we can construct an  $(n, k_\mathrm{v}, k_\mathrm{u}, T_\mathrm{v}, T_\mathrm{u})_{\mathbb{F}}$-streaming code which corrects any $B$-erasure sequence.
\end{Lemma}

\begin{Example}\label{exampleBlockToConv}
Suppose we are given a $(5, 2, 1, 3, 2)_{\mathbb{F}}$-block code which corrects any length-$2$ burst erasure with generator matrix
\begin{align*}
\mathbf{G}=\left[\begin{array}{cccccc}1&0&0& 1&0\\0&1&0&0&1\\0&0&1&1&1 \end{array}\right].
\end{align*}
 Let $\{\mathbf{v}_i\}_{i\in\mathbb{Z}_+}$ and $\{\mathbf{u}_i\}_{i\in\mathbb{Z}_+}$ be the messages of the less-urgent stream and urgent stream respectively where $\mathbf{v}_i=\big[v_i[0] \ v_i[1]\big]\in \mathbb{F}^{2}$ and  $\mathbf{u}_i= u_i[0]\in \mathbb{F}$. From time~$i-2$ to $i+4$, the symbols yielded by the $(5, 2, 1, 3, 2)_{\mathbb{F}}$-streaming code constructed by interleaving the block code according to Lemma~\ref{lemmaBlockToStreaming} are shown in Table~\ref{tableInterleave}.
\begin{table}[!t]
\centering
\begin{tabular}{|c|*{8}{c|}}\hline
\backslashbox{\small Symbol}{\small Time $m$}
&\makebox[1.5em]{$i-2$}&\makebox[1.5em]{$i-1$}&\makebox[1.5em]{$i$}&\makebox[1.5em]{$i+1$}&\makebox[1.5em]{$i+2$}
&\makebox[1.5em]{$i+3$}&\makebox[1.5em]{$i+4$} \\\hline
$x_m[0]=v_m[0]$ &\textcolor{red}{$\boxed{v_{i-2}[0]}$} &\magenta{$v_{i-1}[0]$} & \blue{$v_i[0]$} &$v_{i+1}[0]$&$v_{i+2}[0]$&$v_{i+3}[0]$ &$v_{i+4}[0]$ \\\hline
$x_m[1]=v_m[1]$ & $v_{i-2}[1]$ &$\textcolor{red}{\boxed{v_{i-1}[1]}}$ & $\magenta{v_i[1]}$ & \blue{$v_{i+1}[1]$}&$v_{i+2}[1]$&$v_{i+3}[1]$ &$v_{i+4}[1]$ \\\hline
$x_m[2]=u_m[0]$ &$u_{i-2}[0]$ &$u_{i-1}[0]$ & $\textcolor{red}{\boxed{u_i[0]}}$ &\magenta{$u_{i+1}[0]$}& \blue{$u_{i+2}[0]$}&$u_{i+3}[0]$ &$u_{i+4}[0]$ \\\hline
\parbox[c]{1.2 in}{$x_m[3]=v_{m-3}[0] \\ \text{\qquad\qquad}+u_{m-1}[0]$} &$\ddots$ & $\ddots$& $\ddots$ &\textcolor{red}{\boxed{\parbox[c]{0.5 in}{$v_{i-2}[0]\\+u_i[0]$}}} &\magenta{\parbox[c]{0.6 in}{$v_{i-1}[0]\\+u_{i+1}[0]$}} &  \blue{\parbox[c]{0.6 in}{$v_i[0]\\+u_{i+2}[0]$}} &$\ddots$ \\\hline
\parbox[c]{1.2 in}{$x_m[4]=v_{m-3}[1]\\ \text{\qquad\qquad}+u_{m-2}[0]$} &$\ddots$ &$\ddots$ &$\ddots$  &$\ddots$ &\textcolor{red}{\boxed{\parbox[c]{0.5 in}{$v_{i-1}[1]\\+ u_{i}[0]$}}} &  \parbox[c]{0.6 in}{\magenta{$v_{i}[1]\\+ u_{i+1}[0]$}}&   \parbox[c]{0.6 in}{\blue{$v_{i+1}[1]\\+ u_{i+2}[0]$}} \\\hline
\end{tabular}\smallskip
\caption{Symbols yielded by a $(5, 2, 1, 3, 2)_{\mathbb{F}}$-streaming code through interleaving a $(5, 2, 1, 3, 2)_{\mathbb{F}}$-block code.}
\label{tableInterleave}
\end{table}
 The symbols in Table~\ref{tableInterleave} which are highlighted in the same color diagonally (in~$\searrow$ direction) {are the components of the same codeword with generator matrix $\mathbf{G}$.} Given the fact that the $(5, 2, 1, 3, 2)_{\mathbb{F}}$-block code corrects any length-$2$ burst erasure, we can see from Table~\ref{tableInterleave} that $\big[v_i[0] \ v_i[1]\big]$ and $u_i[0]$ can be perfectly recovered by time $i+3$ and time $i+2$ respectively as long as the erasure sequence is taken from $\Omega_2^5$.
\end{Example}

Lemma~\ref{lemmaBlockToStreaming} reduces the problem of finding high-rate streaming codes which correct any $B$-erasure sequence to the problem of finding high-rate block codes which correct any $B$-erasure sequence. We will construct high-rate block codes by superimposing the codewords of two single-stream block codes, and therefore we need the following definition of a single-stream block code.
\medskip
\begin{Definition}\label{definitionBlockCodeSingleStream}
An $(n, k, 0, T, 0)_{\mathbb{F}}$-block code is also called an \emph{$(n, k, T)_\mathbb{F}$-block code}. The $(n, k, T)_\mathbb{F}$-block code is said to correct a $B$-erasure sequence $e^n$ if the equivalent $(n, k, 0, T, 0)_{\mathbb{F}}$-block code corrects~$e^n$.
\end{Definition}
\medskip

By Definition~\ref{definitionBlockCodeSingleStream}, an $(n, k, T)_\mathbb{F}$-block code ignores the urgent stream of messages by letting the message size for the urgent stream be zero. The following lemma is a restatement of an existing construction~\cite[Th.~2]{MartinianSundberg2004} (see also \cite[Remark~3]{FKLTZA2018}) of an $(n, k, T)_\mathbb{F}$-block code with rate $k/n=\mathrm{C}(T,B)$ which corrects any length-$B$ burst erasure.
\begin{Lemma}\label{lemmaBlockCodeAch}
Suppose $T\ge B\ge 1$ and let $k\triangleq T$ and $n\triangleq k+B$. Fix any $\mathbb{F}$ with $|\mathbb{F}|\ge T$ such that a systematic MDS $(T, T-B)$-code always exists. Let $\mathbf{P}$ denote the parity matrix of the MDS code such that the generator matrix of the MDS code equals $[\mathbf{I}_{T-B}\ \mathbf{P}]$. Then, the $(n, k, T)_\mathbb{F}$-block code with rate $k/n=\mathrm{C}(T,B)$ and generator matrix~$\mathbf{G}$ defined as
\[
\mathbf{G}\triangleq \left[\begin{array}{c:c:c} \mathbf{I}_B & \mathbf{0}^{B\times (T-B)} &  \mathbf{I}_B\\\mathbf{0}^{(T-B)\times B}& \mathbf{I}_{T-B} & \mathbf{P}\end{array}\right]
\]
corrects any length-$B$ burst erasure.
\end{Lemma}

\subsection{Achievability Proof of Theorem~\ref{thmMainResult}}
Fix any $(T_\mathrm{v},T_\mathrm{u},B)$ that satisfies~\eqref{assumptionWholePaper} and~\eqref{assumptionWholePaper*}. Our goal is to show that $\mathcal{C}_{T_\mathrm{v}, T_\mathrm{u},B} \supseteq
\mathcal{R}_{\{T_\mathrm{v}> T_\mathrm{u}+B\}}$ where $\mathcal{R}_{\{T_\mathrm{v}> T_\mathrm{u}+B\}}$ is defined in~\eqref{defSetR} and illustrated in Figure~\ref{figureCapacity}(b). By Corollary~\ref{corollaryConvex}, it suffices to show that the four corner points of $\mathcal{R}_{\{T_\mathrm{v}> T_\mathrm{u}+B\}}$ are $(T_\mathrm{v}, T_\mathrm{u},B)$-achievable. Since the corner points $(0,0)$, $(\mathrm{C}(T_\mathrm{v},B), 0)$ and $(0, \mathrm{C}(T_\mathrm{u},B))$ are $(T_\mathrm{v}, T_\mathrm{u},B)$-achievable by Corollary~\ref{corollaryAch}, it suffices to show that the remaining corner point $\big(\frac{T_\mathrm{v}-T_\mathrm{u}}{T_\mathrm{v}+B},\frac{T_\mathrm{u}}{T_\mathrm{v}+B}\big)$ is $(T_\mathrm{v}, T_\mathrm{u},B)$-achievable. To this end, we let $k_\mathrm{v}\triangleq T_\mathrm{v}-T_\mathrm{u}>0$, $k_\mathrm{u}\triangleq T_\mathrm{u}$ and $n\triangleq T_\mathrm{v}+B$, and will construct an~$(n, k_\mathrm{v}, k_\mathrm{u}, T_\mathrm{v}, T_\mathrm{u})_{\mathbb{F}}$-block code which corrects any length-$B$ burst erasure, which together with Lemma~\ref{lemmaBlockToStreaming} will imply that $\big(\frac{T_\mathrm{v}-T_\mathrm{u}}{T_\mathrm{v}+B},\frac{T_\mathrm{u}}{T_\mathrm{v}+B}\big)$ is $(T_\mathrm{v}, T_\mathrm{u},B)$-achievable. The construction of the $(n, k_\mathrm{v}, k_\mathrm{u}, T_\mathrm{v}, T_\mathrm{u})_{\mathbb{F}}$-block code is described as follows. Fix any~$\mathbb{F}$ with $|\mathbb{F}|\ge \max\{T_\mathrm{u}, T_\mathrm{v}-T_\mathrm{u}\}$. The vectors of less-urgent source symbols and urgent source symbols are denoted by $\vec v=[v[0]\ v[1]\ \ldots v[T_\mathrm{v}-T_\mathrm{u}-1]]$ and $\vec u=[u[0]\ u[1]\ \ldots u[T_\mathrm{u}-1]]$ respectively. Let $\mathbf{V}$ and $\mathbf{U}$ be the parity matrices of a systematic MDS $(T_\mathrm{v}-T_\mathrm{u}, T_\mathrm{v}-T_\mathrm{u}-B)$-code and a systematic MDS $(T_\mathrm{u}, T_\mathrm{u}-B)$-code respectively, and let
\begin{align}
\mathbf{G}\triangleq \left[\begin{array}{c:c:c} \mathbf{I}_{T_\mathrm{v}-T_\mathrm{u}} & \begin{matrix}\mathbf{I}_B \\ \mathbf{V}\end{matrix} & \mathbf{0}^{(T_\mathrm{v}-T_\mathrm{u})\times B} \\\hdashline \mathbf{0}^{T_\mathrm{u}\times(T_\mathrm{v}-T_\mathrm{u})}& \begin{matrix} \mathbf{I}_B\\ \mathbf{0}^{(T_\mathrm{u}-B)\times B} \end{matrix} &  \begin{array}{c:c} \mathbf{0}^{B\times (T_\mathrm{u}-B)} & \mathbf{I}_B\\  \mathbf{I}_{T_\mathrm{u}-B}& \mathbf{U} \end{array}\end{array}\right] \label{defMatrixG}
\end{align}
be the generator matrix of the $(n, k_\mathrm{v}, k_\mathrm{u}, T_\mathrm{v}, T_\mathrm{u})_{\mathbb{F}}$-block code. The intuition behind the construction of $\mathbf{G}$ is to superimpose the codeword generated from the less-urgent symbols
\begin{align}
[x^{(\mathrm{v})}[0]\ x^{(\mathrm{v})}[1]\ \ldots \ x^{(\mathrm{v})}[T_\mathrm{v}-T_\mathrm{u}+B-1]] \triangleq \vec v \left[\begin{array}{c:c}\mathbf{I}_{T_\mathrm{v}-T_\mathrm{u}} & \begin{matrix}\mathbf{I}_B \\ \mathbf{V}\end{matrix}\end{array}\right] \label{defXv}
\end{align}
and the codeword generated from the urgent symbols
\begin{align}
[x^{(\mathrm{u})}[0]\ x^{(\mathrm{u})}[1]\ \ldots \ x^{(\mathrm{u})}[T_\mathrm{u}+B-1]] \triangleq \vec u \left[\begin{array}{c:c}\mathbf{I}_{T_\mathrm{u}} & \begin{matrix}\mathbf{I}_B \\ \mathbf{U}\end{matrix}\end{array}\right] \label{defXu}
\end{align}
such that the two streams interfere with each other in the resultant codeword at~$B$ consecutive positions. Applying Lemma~\ref{lemmaBlockCodeAch} to~\eqref{defXv} and~\eqref{defXu}, we obtain the following two properties for the less-urgent symbols and urgent symbols respectively for each $e^n\in\Omega_B^n$:
\begin{enumerate}
\item[(i)] For each $i\in\{0, 1, \ldots, T_\mathrm{v}-T_\mathrm{u}-1\}$, $v[i]$ can be perfectly recovered from the following set of packets that are not erased by the length-$B$ burst erasure specified by~$e^n$:
\[
\left\{\left. x^{(\mathrm{v})}[\ell]\right| \ell\in\{0, 1, \ldots, \min\{i+T_\mathrm{v}-T_\mathrm{u},T_\mathrm{v}-T_\mathrm{u}+B-1\}\}, e_\ell=0\right\}.
\]

\item[(ii)] For each $i\in\{0, 1, \ldots, T_\mathrm{u}-1\}$, $u[i]$ can be perfectly recovered from the non-erased packets
\[
\left\{\left. x^{(\mathrm{u})}[\ell]\right| \ell\in\{0, 1, \ldots, \min\{i+T_\mathrm{u},T_\mathrm{u}+B-1\}\}, e_\ell=0\right\}.
\]
\end{enumerate}
 Combining~\eqref{defMatrixG}, \eqref{defXv} and~\eqref{defXu}, we obtain
\begin{align}
&[x[0]\ x[1]\ \ldots \ x[n-1]] =\notag\\*
&  \left[\begin{array}{c:c:c} x^{(\mathrm{v})}[0]\ \ldots\ x^{(\mathrm{v})}[T_\mathrm{v}-T_\mathrm{u}-1] & \begin{matrix}x^{(\mathrm{v})}[T_\mathrm{v}-T_\mathrm{u}]\\ +\\ x^{(\mathrm{u})}[0]\end{matrix}\ \ldots\ \begin{matrix}x^{(\mathrm{v})}[T_\mathrm{v}-T_\mathrm{u}+B-1]\\ + \\ x^{(\mathrm{u})}[B-1]\end{matrix} & x^{(\mathrm{u})}[B]\ \ldots \ x^{(\mathrm{u})}[T_\mathrm{u}+B-1] \end{array}\right] \label{defSymbolsX}
\end{align}
where the last~$B$ symbols of the less-urgent stream codeword interfere with the first~$B$ symbols of the urgent stream codeword. In order to show that the~$(n, k_\mathrm{v}, k_\mathrm{u}, T_\mathrm{v}, T_\mathrm{u})_{\mathbb{F}}$-block code defined by~\eqref{defMatrixG} corrects any length-$B$ burst erasure, we fix an arbitrary $e^n\in\Omega_B^n$ and would like to show the following two properties:
\begin{enumerate}
\item[(I)] For each $i\in\{0, 1, \ldots, T_\mathrm{v}-T_\mathrm{u}-1\}$, {suppose the less-urgent symbol $v[i]$ is generated at time~$i$. Then, $v[i]$ can be perfectly recovered with delay~$T_\mathrm{v}$ by time $i+T_\mathrm{v}$ from the following set of packets that are not erased by the length-$B$ burst erasure specified by~$e^n$:
\begin{equation}
\left\{\left. x[\ell]\right| \ell\in\{0, 1, \ldots, \min\{i+T_\mathrm{v},n-1\}\}, e_\ell=0\right\}. \label{eqn_nonerased_packet}
\end{equation}}
\item[(II)] For each $i\in\{0, 1, \ldots, T_\mathrm{u}-1\}$, {suppose the urgent symbol $u[i]$ is generated at time~$T_\mathrm{v}-T_\mathrm{u}+i$. Then, $u[i]$ can be perfectly recovered with delay~$T_\mathrm{u}$ by time $(T_\mathrm{v}-T_\mathrm{u}+i) + T_\mathrm{u} = i+T_\mathrm{v}$ from the set of non-erased packets as stated in~\eqref{eqn_nonerased_packet}.}
\end{enumerate}
We will show Properties~(i) and (ii) in each of the following two cases:\\
\textbf{Case $\{\left.i\in\{0, 1, \ldots, T_\mathrm{v}-T_\mathrm{u}-1\} \right|e_i=1\} =\emptyset$:}\\
By the hypothesis, $\vec v$ can be perfectly recovered by time~$T_\mathrm{v}-T_\mathrm{u}-1$ and hence Property~(I) holds. It remains to prove Property~(II). To this end, we first observe from~\eqref{defSymbolsX} that $x^{(\mathrm{v})}[T_\mathrm{v}-T_\mathrm{u}], x^{(\mathrm{v})}[T_\mathrm{v}-T_\mathrm{u}+1], \ldots, x^{(\mathrm{v})}[T_\mathrm{v}-T_\mathrm{u}+B-1]$ can be perfectly recovered by time~$T_\mathrm{v}-T_\mathrm{u}-1$ because they are functions of~$\vec v$ by~\eqref{defXv}. Therefore, it follows from~\eqref{defSymbolsX} and~\eqref{defXu} that the destination can construct
\[
\left\{\left. x^{(\mathrm{u})}[\ell]\right| \ell\in\{0, 1, \ldots, \min\{i+T_\mathrm{u},T_\mathrm{u}+B-1\}\}, e_\ell=0\right\}
\]
by time~$i+T_\mathrm{v}$ for each $i\in\{0, 1, \ldots, T_\mathrm{u}-1\}$, which implies from the fact $e^n\in\Omega_B^n$ and Property~(ii) that the destination can perfectly recover $x^{(\mathrm{u})}[i]$ by time $i+T_\mathrm{v}$ for each $i\in\{0, 1, \ldots, T_\mathrm{u}-1\}$, and hence Property~(II) holds.
\\
\textbf{Case $\{\left.i\in\{0, 1, \ldots, T_\mathrm{v}-T_\mathrm{u}-1\} \right|e_i=1\} \ne \emptyset$:}\\
In view of the hypothesis and~\eqref{defSymbolsX} and using the fact $e^n\in\Omega_B^n$, we conclude that the destination receives $x^{(\mathrm{u})}[i]$ at time $T_\mathrm{v}-T_\mathrm{u}+i$ for each $i\in\{B, B+1, \ldots, T_\mathrm{u}+B-1\}$ and hence Property~(II) holds. It remains to prove Property~(I). To this end, we first observe from {Property~(ii)} that the destination can perfectly recover $x^{(\mathrm{u})}[i]=u[i]$ by time~$i+T_\mathrm{v}$ for each $i\in\{0, 1, \ldots, T_\mathrm{u}-1\}$. Therefore, it follows from~\eqref{defSymbolsX} that the destination can construct
\[
\left\{\left. x^{(\mathrm{v})}[\ell]\right| \ell\in\{0, 1, \ldots, \min\{i+T_\mathrm{v},T_\mathrm{v}-T_\mathrm{u}+B-1\}\}, e_\ell=0\right\}
\]
by time~$i+T_\mathrm{v}$ for each $i\in\{0, 1, \ldots, T_\mathrm{v}-T_\mathrm{u}+B-1\}$, which implies from the fact $e^n\in\Omega_B^n$ and Property~(i) that the destination can perfectly recover $x^{(\mathrm{v})}[i]$ by time $i+T_\mathrm{v}$ for each $i\in\{0, 1, \ldots, T_\mathrm{v}-T_\mathrm{u}-1\}$, and hence Property~(I) holds.

Combining the above two cases, we conclude the Properties~(I) and (II) hold for all $e^n\in\Omega_B^n$, which implies that the~$(n, k_\mathrm{v}, k_\mathrm{u}, T_\mathrm{v}, T_\mathrm{u})_{\mathbb{F}}$-block code defined by~\eqref{defMatrixG} corrects any length-$B$ burst erasure,  which together with Lemma~\ref{lemmaBlockToStreaming} implies that $\big(\frac{T_\mathrm{v}-T_\mathrm{u}}{T_\mathrm{v}+B},\frac{T_\mathrm{u}}{T_\mathrm{v}+B}\big)$ is $(T_\mathrm{v}, T_\mathrm{u},B)$-achievable (cf.\ Definition~\ref{definitionAchievability}).
\section{Converse Proof of Main Result} \label{sectionConverse}
Our goal is to show that
$
\mathcal{C}_{T_\mathrm{v}, T_\mathrm{u},B} \subseteq
\mathcal{R}_{\{T_\mathrm{v}> T_\mathrm{u}+B\}}
$.
Equivalently, we would like to show that~\eqref{capacityEq2} and~\eqref{capacityEq3} hold.
To this end, we let $(R_\mathrm{v}, R_\mathrm{u})$ be a rate pair in $\mathcal{C}_{T_\mathrm{v}, T_\mathrm{u},B}$. Fix an arbitrary~$\delta>0$. By Definition~\ref{definitionAchievability} and Definition~\ref{definitionCapacity}, there exists an $(n, k_\mathrm{v}, k_\mathrm{u}, T_\mathrm{v}, T_\mathrm{u})_{\mathbb{F}}$-streaming code which corrects any $B$-erasure sequence such that
\begin{align}
\frac{k_\mathrm{v}}{n}\ge R_\mathrm{v}-\delta \label{converseProofRateRv}
\end{align}
 and
 \begin{align}
 \frac{k_\mathrm{u}}{n}\ge R_\mathrm{u}-\delta. \label{converseProofRateRu}
 \end{align}
\subsection{Sum-Rate Bound~\eqref{capacityEq2}}
By Definition~\ref{definitionCode}, the $(n, k_\mathrm{v}, k_\mathrm{u}, T_\mathrm{v}, T_\mathrm{u})_{\mathbb{F}}$-streaming code can be viewed as an $(n, k_\mathrm{v}+k_\mathrm{u}, 0, T_\mathrm{v}, 0)_{\mathbb{F}}$-streaming code which corrects any $B$-erasure sequence. Consequently, the sum-rate for the $(n, k_\mathrm{v}, k_\mathrm{u}, T_\mathrm{v}, T_\mathrm{u})_{\mathbb{F}}$-streaming codes must not exceed the single-stream capacity $\mathrm{C}(T_\mathrm{v}, B)$ (cf.\ Section~\ref{subsecKnownResults}), which implies that
\begin{align}
\frac{k_\mathrm{v}+k_\mathrm{u}}{n} \le \mathrm{C}(T_\mathrm{v}, B). \label{eq1ConverseProof}
\end{align}
Combining~\eqref{eq1ConverseProof}, \eqref{converseProofRateRv} and \eqref{converseProofRateRu}, we have
\begin{align*}
R_\mathrm{u} + R_\mathrm{v} \le \mathrm{C}(T_\mathrm{v}, B) + 2\delta,
\end{align*}
which then implies~\eqref{capacityEq2} by taking the limit $\delta\to 0$.
\subsection{Genie-Aided Bound~\eqref{capacityEq3}}
\begin{figure}[!t]
\centering
   \includegraphics[width=4 in]{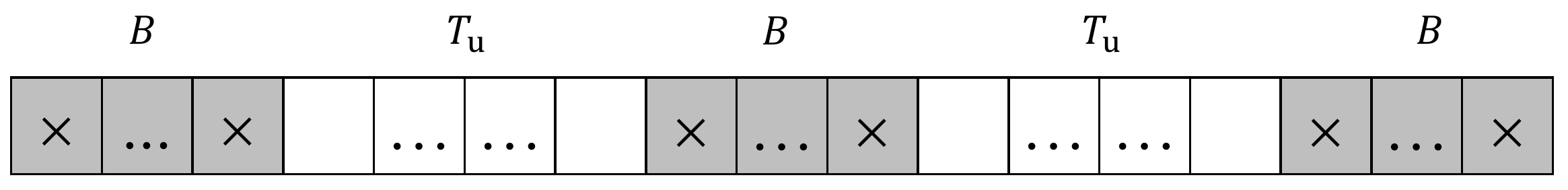}
\caption{The periodic erasure sequence $\boldsymbol{\varepsilon}^{(0)}$.}
\label{figureErasurePattern}
\end{figure}
{Given the $(n, k_\mathrm{v}, k_\mathrm{u}, T_\mathrm{v}, T_\mathrm{u})_{\mathbb{F}}$-streaming code that corrects any $B$-erasure sequence and satisfies~\eqref{converseProofRateRv} and~\eqref{converseProofRateRu}, we use Corollary~\ref{corollaryConcatenate} to construct for each $q\in\mathbb{N}$ a $(qn, qk_\mathrm{v}, qk_\mathrm{u}, T_\mathrm{v}, T_\mathrm{u})_{\mathbb{F}}$-streaming code that corrects any $B$-erasure sequence and satisfies~\eqref{converseProofRateRv} and~\eqref{converseProofRateRu}.
Fix any $q\in\mathbb{N}$.
 In order to develop a genie-aided bound associated with the $(qn, qk_\mathrm{v}, qk_\mathrm{u}, T_\mathrm{v}, T_\mathrm{u})_{\mathbb{F}}$-streaming code, we let~$\mathbf{u}_0$ and~$\mathbf{v}_0$ be the urgent and less-urgent source packets which are uniformly distributed on $\mathbb{F}^{qk_\mathrm{u}}$ and $\mathbb{F}^{qk_\mathrm{v}}$ respectively, and assume that $\{(\mathbf{u}_i,\mathbf{v}_i)\}_{i\in\mathbb{Z}_+}$ are independent and identically distributed (i.i.d.).}
{The genie-aided bound associated with the $(qn, qk_\mathrm{v}, qk_\mathrm{u}, T_\mathrm{v}, T_\mathrm{u})_{\mathbb{F}}$-streaming code is obtained by considering the following set of periodic erasure patterns:} Define
\begin{align*}
n_\mathrm{u}\triangleq T_\mathrm{u}+B
\end{align*}
and construct for each $\Delta\in\{0, 1, \ldots, T_\mathrm{u}+B-1\}$ a periodic erasure pattern $\boldsymbol{\varepsilon}^{(\Delta)}=\{\varepsilon_i^{(\Delta)}\}_{i=0}^\infty$ as
\begin{align}
\varepsilon_i^{(\Delta)} \triangleq
\begin{cases}
1 & \text{if $i-\Delta\in \{\kappa n_\mathrm{u}, \kappa n_\mathrm{u}+1, \ldots,   \kappa n_\mathrm{u} + B-1\}$ for some $\kappa\in\mathbb{Z}$,} \\
0 & \text{otherwise.}
\end{cases} \label{defVarepsilon}
\end{align}
In other words, we construct $\boldsymbol{\varepsilon}^{(\Delta)}$ by offsetting the periodic erasure sequence $\boldsymbol{\varepsilon}^{(0)}$ by~$\Delta$ time slots to the right, where $\boldsymbol{\varepsilon}^{(0)}$ is illustrated in Figure~\ref{figureErasurePattern}.
By construction, each $\boldsymbol{\varepsilon}^{(\Delta)}$ has a period of $n_\mathrm{u}$ time slots and each period consists of an initial length-$B$ burst erasure followed by a length-$T_\mathrm{u}$ noiseless duration.
{Fix a erasure pattern $\boldsymbol{\varepsilon}^{(\Delta)}$, and we will obtain a corresponding genie-aided bound associated with the $(qn, qk_\mathrm{v}, qk_\mathrm{u}, T_\mathrm{v}, T_\mathrm{u})_{\mathbb{F}}$-streaming code in the following.} Let
\begin{align}
\beta_m
= H\big(\mathbf{x}_m\big| \mathbf{u}^\infty, \{\mathbf{x}_\ell\}_{\ell=0}^{m-1}\big) \label{eq5ConverseProof}
\end{align}
be the conditional entropy of~$\mathbf{x}_m$ conditioned on $(\mathbf{u}^\infty, \{\mathbf{x}_\ell\}_{\ell=0}^{m-1})$ for each~$m\in \{0, 1, \ldots, qn_\mathrm{u}-1\}$, which specifies the approximate number of bits required to construct~$\mathbf{x}_m$ based on the knowledge of $(\mathbf{u}^\infty, \{\mathbf{x}_\ell\}_{\ell=0}^{m-1})$. Suppose we use a standard arithmetic code~\cite[Ch.~13.3]{Cover06} to compress $\mathbf{x}_m$ conditioned on each outcome assumed by $(\mathbf{u}^\infty, \{\mathbf{x}_\ell\}_{\ell=0}^{m-1})$ for each $m\in\{0, 1, \ldots, qn_\mathrm{u}-1\}$, and let $\hat X_m$  be the noiseless compressed version of $\mathbf{x}_m$ {accessible to the genie} that satisfies the equality
\begin{align}
H\big(\mathbf{x}_m\big| \mathbf{u}^\infty, \{\mathbf{x}_\ell\}_{\ell=0}^{m-1},\hat X_m\big)=0 \label{eq6ConverseProof}
\end{align}
and the inequality (cf.\ \cite[Ch.~13.3]{Cover06})
\begin{align}
H(\hat X_m)
&\le H\big(\mathbf{x}_m\big| \mathbf{u}^\infty, \{\mathbf{x}_\ell\}_{\ell=0}^{m-1}\big)  + 2 \notag \\*
& = \beta_m + 2 \label{eq7ConverseProof}
\end{align}
where the last equality is due to~\eqref{eq5ConverseProof}.
{According to the arithmetic code,} the random variable $\hat X_m$ is constructed by first generating $\{\mathbf{u}_\ell\}_{\ell=0}^{m}$ and $\{\mathbf{x}_\ell\}_{\ell=0}^{m-1}$ followed by generating $\hat X_m$ based on the conditional distribution $p_{\hat X_m|\{\mathbf{u}_\ell\}_{\ell=0}^{m},\{\mathbf{x}_\ell\}_{\ell=0}^{m-1}}$. {Note that for the special case where the $(qn, qk_\mathrm{v}, qk_\mathrm{u}, T_\mathrm{v}, T_\mathrm{u})_{\mathbb{F}}$-streaming code is systematic, it can be seen that $\{\mathbf{x}_\ell\}_{\ell=0}^{m-1}$ contains $\{\mathbf{v}_\ell\}_{\ell=0}^{m-1}$ and hence setting $\hat X_m$ equal to $\mathbf{v}_m$ suffices to yield~\eqref{eq6ConverseProof} and~\eqref{eq7ConverseProof}. However, for the general case where the streaming code can be non-systematic, the arithmetic coding argument is needed for obtaining~\eqref{eq6ConverseProof} and~\eqref{eq7ConverseProof}.}
 In order to obtain the genie-aided bound corresponding to the fixed~$q$, the fixed~$(qn, qk_\mathrm{v}, qk_\mathrm{u}, T_\mathrm{v}, T_\mathrm{u})_{\mathbb{F}}$-streaming code and the fixed~$\boldsymbol{\varepsilon}^{(\Delta)}$, we suppose the genie provides the destination with
$\{\hat X_m\,|\,m\in\mathcal{A}_{\Delta}\}$
where
  \begin{equation}
 \mathcal{A}_{\Delta}\triangleq \big\{i\in\mathbb{Z}_+\big|  \varepsilon_i^{(\Delta)}=1 \big\} \label{defSetA}
 \end{equation}
 {denotes the set of time indices at which the transmitted packets are erased according to~$\boldsymbol{\varepsilon}^{(\Delta)}$.}
 To simplify notation, we let $\mathcal{A}_{\Delta}^c \triangleq \mathbb{Z}_+\setminus  \mathcal{A}_{\Delta}$.
 Then we claim that every urgent source packet and every less-urgent source packet can be recovered when the erasure sequence is $\boldsymbol{\varepsilon}^{(\Delta)}$ (cf.\ \eqref{defVarepsilon}). To prove the claim, we consider the following chain of inequalities:
\begin{align}
&H\big(\{(\mathbf{u}_i, \mathbf{v}_{i})\}_{i=0}^{q n_\mathrm{u}-T_\mathrm{u}-T_\mathrm{v}-1}\big|\{\mathbf{x}_m: \mathcal{A}_{\Delta}^c\cap[0, q n_\mathrm{u}-1]\}, \{\hat X_m: m \in \mathcal{A}_{\Delta}\cap[0,q n_\mathrm{u}-1]\} \big) \notag\\*
&\le H\big(\{\mathbf{x}_i\}_{i=0}^{q n_\mathrm{u}-T_\mathrm{u}-1}\big|\{\mathbf{x}_m: \mathcal{A}_{\Delta}^c\cap[0, q n_\mathrm{u}-1]\}, \{\hat X_m: m \in \mathcal{A}_{\Delta}\cap[0,q n_\mathrm{u}-1]\} \big)\label{eq8ConverseProofa}  \\
&= \sum_{i=0}^{q n_\mathrm{u}-T_\mathrm{u}-1} H\big(\mathbf{x}_i\big|\{\mathbf{x}_\ell\}_{\ell=0}^{i-1}, \{\mathbf{x}_m: m\in\mathcal{A}_{\Delta}^c\cap [i, q n_\mathrm{u}-1]\}, \{\hat X_m: m \in \mathcal{A}_{\Delta}\cap[0,q n_\mathrm{u}-1]\}\big) \notag\\
&\le \sum_{i=0}^{q n_\mathrm{u}-T_\mathrm{u}-1} H\big(\mathbf{x}_i\big|\{\mathbf{x}_\ell\}_{\ell=0}^{i-1}, \{\mathbf{x}_m: m\in\mathcal{A}_{\Delta}^c\cap [i, i+T_\mathrm{u}]\}, \{\hat X_m: m \in \mathcal{A}_{\Delta}\cap[0,q n_\mathrm{u}-1]\}\big)  \notag\\
&= \sum_{i=0}^{q n_\mathrm{u}-T_\mathrm{u}-1} \!\!\!\!\! H\big(\mathbf{x}_i\big|\{\mathbf{u}_\ell\}_{\ell=0}^{i},\{\mathbf{x}_\ell\}_{\ell=0}^{i-1},\{\mathbf{x}_m: m\in\mathcal{A}_{\Delta}^c\cap[i, i+T_\mathrm{u}]\},  \{\hat X_m: m \in \mathcal{A}_{\Delta}\cap[0,q n_\mathrm{u}-1]\}\big) \label{eq8ConverseProofb}\\
&\le \sum_{i=0}^{q n_\mathrm{u}-T_\mathrm{u}-1} \mathbf{1}\{\varepsilon_i^{(\Delta)}=1\}\times H\big(\mathbf{x}_i\big| \{ \mathbf{u}_\ell\}_{\ell=0}^i, \{\mathbf{x}_\ell\}_{\ell=0}^{i-1}, \hat X_i \big) \label{eq8ConverseProofc}\\
&= 0\label{eq8ConverseProof}
\end{align}
where
\begin{itemize}
\item \eqref{eq8ConverseProofa} is due to that fact that $\{(\mathbf{u}_i, \mathbf{v}_{i})\}_{i=0}^{q n_\mathrm{u}-T_\mathrm{u}-T_\mathrm{v}-1}$ is a function of~$\{\mathbf{x}_i\}_{i=0}^{qn_\mathrm{u}-T_\mathrm{u}-1}$.
\item \eqref{eq8ConverseProofb} is due to the fact that $\{\mathbf{u}_\ell\}_{\ell=0}^{i}$ is a function of $(\{\mathbf{x}_\ell\}_{\ell=0}^{i-1},\{\mathbf{x}_m: m\in\mathcal{A}_{\Delta}^c\cap[i, i+T_\mathrm{u}]\})$, which is a direct consequence of the fact that the $(qn, qk_\mathrm{v}, qk_\mathrm{u}, T_\mathrm{v}, T_\mathrm{u})_{\mathbb{F}}$-streaming code corrects any $B$-erasure sequence.
\item \eqref{eq8ConverseProofc} is due to the definition of~$\mathcal{A}_{\Delta}$ in~\eqref{defSetA}.
\item \eqref{eq8ConverseProof} is due to~\eqref{eq6ConverseProof}.
\end{itemize}
{Equation~\eqref{eq8ConverseProof} implies that the urgent and less-urgent source packets generated before time~$q n_\mathrm{u}-T_\mathrm{u}-T_\mathrm{v}-1$ can be recovered by the destination by time $q n_\mathrm{u}+1$ if the erasure sequence is $\boldsymbol{\varepsilon}^{(\Delta)}$ and the genie provides the destination with the side information $\{\hat X_m: m \in \mathcal{A}_{\Delta}\cap[0,q n_\mathrm{u}-1]\}$.}
Therefore, it follows from~\eqref{eq8ConverseProof} and~\eqref{eq7ConverseProof} that
\begin{align}
q(k_\mathrm{u}+k_\mathrm{v})(q n_\mathrm{u}-T_\mathrm{u}-T_\mathrm{v}) & = H\big(\{(\mathbf{u}_i, \mathbf{v}_{i})\}_{i=0}^{q n_\mathrm{u}-T_\mathrm{u}-T_\mathrm{v}-1}\big) \notag\\
& \le  H\big(\{\mathbf{x}_m: \mathcal{A}_{\Delta}^c\cap[0, q n_\mathrm{u}-1]\}, \{\hat X_m: m \in \mathcal{A}_{\Delta}\cap[0,q n_\mathrm{u}-1]\}\big) \notag\\
& \le H\big(\{\mathbf{x}_m: \mathcal{A}_{\Delta}^c\cap[0, q n_\mathrm{u}-1]\}\big) + H\big(\{\hat X_m: m \in \mathcal{A}_{\Delta}\cap[0,q n_\mathrm{u}-1]\}\big)\notag\\
& \le \sum_{ m \in \mathcal{A}_{\Delta}^c\cap[0,q n_\mathrm{u}-1]} H(\mathbf{x}_m) + \sum_{ m \in \mathcal{A}_{\Delta}\cap[0,q n_\mathrm{u}-1]}(\beta_m+2). \label{eq9ConverseProof}
\end{align}
{
Taking average on both sides of~\eqref{eq9ConverseProof} over $\Delta\in\{0, 1, \ldots, T_\mathrm{u}+B-1\}$, we obtain
\begin{align*}
q(k_\mathrm{u}+k_\mathrm{v})(q n_\mathrm{u}-T_\mathrm{u}-T_\mathrm{v}) \le \frac{1}{T_\mathrm{u}+B}\sum_{\Delta=0}^{T_\mathrm{u}+B-1}\bigg( \sum_{ m \in \mathcal{A}_{\Delta}^c\cap[0,q n_\mathrm{u}-1]} H(\mathbf{x}_m) + \sum_{ m \in \mathcal{A}_{\Delta}\cap[0,q n_\mathrm{u}-1]}(\beta_m+2)\bigg),
\end{align*}
which together with the definition of~$\mathcal{A}_\Delta$ in~\eqref{defSetA} implies that
\begin{align}
q(k_\mathrm{u}+k_\mathrm{v})(q n_\mathrm{u}-T_\mathrm{u}-T_\mathrm{v}) \le \frac{T_\mathrm{u}}{T_\mathrm{u}+B}\sum_{m=0}^{q n_\mathrm{u}-1} H(\mathbf{x}_m) +  \frac{B}{T_\mathrm{u}+B}\sum_{m=0}^{q n_\mathrm{u}-1}(\beta_m+2). \label{eq9ConverseProof+}
\end{align}
Since $H(\mathbf{x}_m) \le qn$ for each $m$ by construction and
\begin{align*}
\sum_{m=0}^{q n_\mathrm{u}-1}\beta_m &=   H\big(\{\mathbf{x}_\ell\}_{\ell=0}^{q n_\mathrm{u}-1}\big| \mathbf{u}^\infty \big)\notag\\*
&\le H\big(\{\mathbf{v}_\ell\}_{\ell=0}^{q n_\mathrm{u}-1}\big) \notag\\*
& \le q^2 n_\mathrm{u}k_\mathrm{v}
\end{align*}
due to~\eqref{eq5ConverseProof} and the fact that $\{\mathbf{x}_\ell\}_{\ell=0}^{q n_\mathrm{u}-1}$ is a function of $\{\mathbf{u}_\ell,\mathbf{v}_\ell\}_{\ell=0}^{q n_\mathrm{u}-1}$, it follows from~\eqref{eq9ConverseProof+} that
\begin{align}
q(k_\mathrm{u}+k_\mathrm{v})(q n_\mathrm{u}-T_\mathrm{u}-T_\mathrm{v}) \le
  \frac{q^2 n n_\mathrm{u}T_\mathrm{u}}{T_\mathrm{u}+B}+  \frac{ q^2 n_\mathrm{u}k_\mathrm{v}B}{T_\mathrm{u}+B}+  \frac{2qn_\mathrm{u}B}{T_\mathrm{u}+B}. \label{eq10ConverseProof}
\end{align}
}
Dividing both sides of~\eqref{eq10ConverseProof} by~$q^2nn_\mathrm{u}$, we obtain
\begin{align*}
\left(\frac{k_\mathrm{u}+k_\mathrm{v}}{n}\right)\left(1-\frac{T_\mathrm{u}+T_\mathrm{v}}{q\mathrm{n}_\mathrm{u}}\right) \le \frac{T_\mathrm{u}}{T_\mathrm{u}+B}+ \frac{B}{T_\mathrm{u}+B}\times\frac{k_\mathrm{v}}{n}+\frac{2B}{qn(T_\mathrm{u}+B)}, 
\end{align*}
which together with the fact $n_\mathrm{u}=T_\mathrm{u}+B$ implies that
\begin{align}
\left(1-\frac{T_\mathrm{u}+T_\mathrm{v}}{q\mathrm{n}_\mathrm{u}}\right)\frac{k_\mathrm{u}}{n} +\left(\frac{T_\mathrm{u}}{n_\mathrm{u}}-\frac{T_\mathrm{u}+T_\mathrm{v}}{q\mathrm{n}_\mathrm{u}}\right)\frac{k_\mathrm{v}}{n} \le \frac{T_\mathrm{u}}{n_\mathrm{u}}+\frac{2B}{qnn_\mathrm{u}}. \label{eq11ConverseProof}
\end{align}
Combining~\eqref{eq11ConverseProof}, the definition of~$\mathrm{C}(\cdot, \cdot)$ in~\eqref{defCapacityPTP}, \eqref{converseProofRateRv} and~\eqref{converseProofRateRu}, we obtain
\begin{align}
\left(1-\frac{T_\mathrm{u}+T_\mathrm{v}}{q\mathrm{n}_\mathrm{u}}\right)(R_\mathrm{u}-\delta) +\left(\mathrm{C}(T_\mathrm{u}, B)-\frac{T_\mathrm{u}+T_\mathrm{v}}{q\mathrm{n}_\mathrm{u}}\right)(R_\mathrm{v}-\delta) \le \mathrm{C}(T_\mathrm{u}, B)+\frac{2B}{qnn_\mathrm{u}}.\label{eq12ConverseProof}
\end{align}
Taking the limit $q\to\infty$ followed by letting $\delta\to 0$ on both sides of~\eqref{eq12ConverseProof}, we obtain~\eqref{capacityEq3}.
\section{Concluding Remarks} \label{conclusion}
We have investigated streaming codes that multiplex an urgent stream of messages with delay constraint~$T_\mathrm{u}$ and a less-urgent stream of messages with delay constraint~$T_\mathrm{v}$ over the deterministic burst-erasure model where $T_\mathrm{v}\ge T_\mathrm{u}$. The capacity region has been proved for the case $T_\mathrm{v}> T_\mathrm{u}+B$ under assumption~\ref{assumptionWholePaper}, which together with the existing results described in Section~\ref{subsecKnownResults} implies the full characterization of the capacity region for all parameters of $(T_\mathrm{v}, T_\mathrm{u},B)$. In particular, the capacity regions for the case $T_\mathrm{u}<T_\mathrm{v}\le T_\mathrm{u} +B$ and the case $T_\mathrm{v}> T_\mathrm{u} +B$ are shown in Figure~\ref{figureCapacity}(a) and Figure~\ref{figureCapacity}(b) respectively. While systematic streaming codes alone achieve the capacity region for the case~$T_\mathrm{u}< T_\mathrm{v}\le T_\mathrm{u} +B$ and the case $T_\mathrm{v}\ge T_\mathrm{u} +2B$ by~\cite[Th.~3]{BLKTZA2018} and Remark~\ref{remark2} respectively, it remains open whether systematic streaming codes are sufficient to achieve the capacity region for the case $T_\mathrm{u}+B < T_\mathrm{v} < T_\mathrm{u}+2B$.

The main result in this paper, i.e., Theorem~\ref{thmMainResult}, is readily generalized to the following deterministic model that generates multiple burst erasures as explained in~{\cite[Remark 1]{EKT2014}} (see also {\cite[Sec.~II]{BLKTZA2018}}): The channel introduces multiple burst erasures on the discrete timeline where the length of each burst does not exceed~$B$ and the length of the guard space between two adjacent bursts is at least~$T_\mathrm{v}$. Future work may generalize the main result to the erasure model which introduces both burst and arbitrary erasures as investigated in~\cite{BPKTA17} and~\cite{FKLTZA2018}.
\appendices

\section{Proof of Corollary~\ref{corollaryConcatenate}} \label{appendixA-}
{Fix an $(n, k_\mathrm{v}, k_\mathrm{u}, T_\mathrm{v}, T_\mathrm{u})_{\mathbb{F}}$-streaming code that corrects any $B$-erasure sequence and fix any $q\in\mathbb{N}$. Construct $q$ instances of the $(n, k_\mathrm{v}, k_\mathrm{u}, T_\mathrm{v}, T_\mathrm{u})_{\mathbb{F}}$-streaming code. Recalling Definition~\ref{definitionCode}, we concatenate the length-$n$ transmitted packets generated at time~$i$ by the $q$ instances of the streaming code and construct at time~$i$ a length-$(qn)$ transmitted packet for each~$i\in\mathbb{Z}_+$. Due to Definition~\ref{definitionCode} and Definition~\ref{definitionRecover}, the concatenated code associated with the sequence of length-$(qn)$ transmitted packets can be viewed as a $(qn, qk_\mathrm{v}, qk_\mathrm{u}, T_\mathrm{v}, T_\mathrm{u})_{\mathbb{F}}$-streaming code which corrects any $B$-erasure sequence.
}

\section{Proof of Corollary~\ref{corollaryConvex}} \label{appendixA}
By Definition~\ref{definitionAchievability} and Definition~\ref{definitionCapacity}, it suffices to prove the following: For any $(n^{(0)}, k_\mathrm{v}^{(0)}, k_\mathrm{u}^{(0)}, T_\mathrm{v}, T_\mathrm{u})_{\mathbb{F}}$-streaming code and any $(n^{(1)}, k_\mathrm{v}^{(1)}, k_\mathrm{u}^{(1)}, T_\mathrm{v}, T_\mathrm{u})_{\mathbb{F}}$-streaming code which correct any $B$-erasure sequence, there exists an $(n, k_\mathrm{v}, k_\mathrm{u}, T_\mathrm{v}, T_\mathrm{u})_{\mathbb{F}}$-streaming code which corrects any $B$-erasure sequence where
\begin{align}
\frac{k_\mathrm{v}}{n} = \frac{k_\mathrm{v}^{(0)}}{2n^{(0)}} + \frac{ k_\mathrm{v}^{(1)}}{2n^{(1)}}\label{eq1AppendixA}
\end{align}
and
\begin{align}
\frac{k_\mathrm{u}}{n} = \frac{k_\mathrm{u}^{(0)}}{2n^{(0)}} + \frac{ k_\mathrm{u}^{(1)}}{2n^{(1)}}. \label{eq2AppendixA}
\end{align}
In order to show~\eqref{eq1AppendixA} and~\eqref{eq2AppendixA}, {we concatenate $n^{(1)}$ instances of the length-$n^{(0)}$ transmitted packet generated at time~$i$ and $n^{(0)}$ instances of the length-$n^{(1)}$ transmitted packet generated at time~$i$ and form at time~$i$ a new length-$(2n^{(0)}n^{(1)})$ transmitted packet for each~$i\in\mathbb{Z}_+$.} By construction, the concatenated code associated with the sequence of length-$(2n^{(0)}n^{(1)})$ transmitted packets can be viewed as a $(n, k_\mathrm{v}, k_\mathrm{u}, T_\mathrm{v}, T_\mathrm{u})_{\mathbb{F}}$-streaming code which corrects any $B$-erasure sequence where $n=2n^{(0)}n^{(1)}$, $k_\mathrm{v}=n^{(1)}k_\mathrm{v}^{(0)} + n^{(0)}k_\mathrm{v}^{(1)}$ and $k_\mathrm{u}=n^{(1)}k_\mathrm{u}^{(0)} + n^{(0)}k_\mathrm{u}^{(1)}$. In particular, the concatenated code satisfies~\eqref{eq1AppendixA} and~\eqref{eq2AppendixA}.

\section{Proof of Lemma~\ref{lemmaBlockToStreaming}} \label{appendixB}
Suppose we are given an $(n, k_\mathrm{v}, k_\mathrm{u}, T_\mathrm{v}, T_\mathrm{u})_{\mathbb{F}}$-block code which corrects any $B$-erasure sequence, and let
$\mathbf{G}\in\mathbb{F}^{(k_\mathrm{v}+ k_\mathrm{u})\times n}$
 be the generator matrix. By Definition~\ref{definitionCodeBlock}, the block code has the following properties:
\begin{enumerate}
\item[(i)] The length of the block code is~$n$.
\item[(ii)] From time $0$ to $n-1$, the symbols
\begin{align*}
\big[x[0] \ x[1]\ \cdots \ x[n-1]\big]=\big[\vec v \ \vec u\big]\mathbf{G}
\end{align*}
     are transmitted.
     \item[(iii)] Upon receiving
 \begin{align*}
 &\big[y[0]\ \ldots\ y[\min\{\ell+T_\mathrm{v},n-1\}]\big] \notag\\
 &\quad=  \big[g_1(x[0], e_0)\ \ldots\ g_1(x[\min\{\ell+T_\mathrm{v},n-1\}], e_{\min\{\ell+T_\mathrm{v},n-1\}})\big],
 \end{align*}
  the destination can perfectly recover $v[\ell]$ by time $\min\{\ell+T_\mathrm{v},n-1\}$ for each $\ell\in\{0, 1, \ldots, k_\mathrm{v}-1\}$ as long as $e^n\in\Omega_B^n$.
 \item[(iv)] Upon receiving
 \begin{align*}
 &\big[y[0]\ \ldots\ y[\min\{\ell+T_\mathrm{u},n-1\}]\big] \notag\\
 &\quad=  \big[g_1(x[0], e_0)\ \ldots\ g_1(x[\min\{\ell+T_\mathrm{u},n-1\}], e_{\min\{\ell+T_\mathrm{u},n-1\}})\big],
 \end{align*}
  the destination can perfectly recover $u[\ell]$ by time $\min\{\ell+T_\mathrm{u},n-1\}$ for each $\ell\in\{0, 1, \ldots, k-1\}$ as long as $e^n\in\Omega_B^n$.
\end{enumerate}

In order to construct $(n, k_\mathrm{v}, k_\mathrm{u}, T_\mathrm{v}, T_\mathrm{u})_{\mathbb{F}}$-streaming code (cf.\ Definition~\ref{definitionCode}) which corrects any length-$B$ burst erasure, we first let $\{\mathbf{v}_i\}_{i=0}^\infty$ denote a sequence of length-$k_\mathrm{v}$ less-urgent packets
and let $\{\mathbf{u}_i\}_{i=0}^\infty$ denote a sequence of length-$k_\mathrm{u}$ urgent packets, and let $v_i[\ell]$ and $u_i[\ell]$ denote the $(\ell+1)^{\text{th}}$ element of $\mathbf{v}_i$ and $\mathbf{u}_i$ respectively such that
\begin{align}
\mathbf{v}_i \triangleq [v_i[0]\ v_i[1]\ \cdots \ v_i[k_\mathrm{v}-1] ] \label{defViAppendixB}
\end{align}
and
\begin{align}
\mathbf{u}_i \triangleq [u_i[0]\ u_i[1]\ \cdots \ u_i[k_\mathrm{u}-1] ] \label{defUiAppendixB}
\end{align}
for all $i\in\mathbb{Z}_+$. Using the convention that $\big[\mathbf{u}_j\ \mathbf{v}_j\big] \triangleq \mathbf{0}^{1\times(k_\mathrm{v}+k_\mathrm{u})}$ for any $j<0$, we construct
\begin{align}
\big[x_i[0]\ x_{i+1}[1]\ \cdots \ x_{i+n-1}[n-1] \big] \triangleq \big[\vec v_i \ \vec u_i \big] \mathbf{G} \label{appendixBeq1}
\end{align}
for each $i\in\{-n+1, -n+2, \ldots\}$ where
\begin{align}
\vec v_i \triangleq \big[v_i[0]\ v_{i+1}[1]\ \ldots \ v_{i+k_\mathrm{v}-1}[k_\mathrm{v}-1] \big], \label{defViAppendixB}
\end{align}
\begin{align}
\vec u_i \triangleq \big[u_{i+k_\mathrm{v}}[0]\ v_{i+k_\mathrm{v}+1}[1]\ \ldots \ v_{i+k_\mathrm{v}+k_\mathrm{u}-1}[k_\mathrm{u}-1] \big], \label{defUiAppendixB}
\end{align}
and $\mathbf{G}$ is the generator matrix of the $(n, k_\mathrm{v}, k_\mathrm{u}, T_\mathrm{v}, T_\mathrm{u})_{\mathbb{F}}$-block code which corrects any length-$B$ burst erasure. In other words, {we are coding $\{[\mathbf{v}_i\ \mathbf{u}_i]: i\in\mathbb{Z}_+\}$ diagonally as illustrated in Table~\ref{tableInterleave} where $x_m[\ell]$ denotes the symbol~$\ell$ transmitted at time~$m$}.
At each time $i\in\mathbb{Z}_+$, the source transmits
\begin{align}
\mathbf{x}_i\triangleq \big[x_i[0]\ x_i[1]\ \cdots \ x_i[n-1]\big]. \label{defXiAppendixB}
\end{align}
 Based on the $(n, k_\mathrm{v}, k_\mathrm{u}, T_\mathrm{v}, T_\mathrm{u})_{\mathbb{F}}$-block code which satisfies Properties~(i) to~(iv) as stated at the beginning of this proof, we have constructed an $(n, k_\mathrm{v}, k_\mathrm{u}, T_\mathrm{v}, T_\mathrm{u})_{\mathbb{F}}$-streaming code where $\mathbf{v}_i$, $\mathbf{u}_i$ and $\mathbf{x}_i$ satisfy~\eqref{appendixBeq1}, \eqref{defViAppendixB}, \eqref{defUiAppendixB} and~\eqref{defXiAppendixB}. Our remaining goal is to show that the $(n, k_\mathrm{v}, k_\mathrm{u}, T_\mathrm{v}, T_\mathrm{u})_{\mathbb{F}}$-streaming code corrects any length-$B$ burst erasure. To this end, we fix any $i\in\mathbb{Z}_+$ and any $\mathbf{e}\in\Omega_B$, and would like to show that the destination can perfectly recover $\mathbf{v}_i=\big[v_i[0]\ v_i[1]\ \cdots \ v_i[k_\mathrm{v}-1]\big]$ based on
  \begin{align}
 [\mathbf{y}_0\ \mathbf{y}_1\ \ldots\ \mathbf{y}_{i+T_\mathrm{v}}]=  [g_n(\mathbf{x}_0, e_0)\ g_n(\mathbf{x}_1, e_1)\ \ldots\ g_n(\mathbf{x}_{i+T_\mathrm{v}}, e_{i+T_\mathrm{v}})],\label{appendixBeq5}
 \end{align}
 and can perfectly recover
 $\mathbf{u}_i=\big[u_i[0]\ u_i[1]\ \cdots \ u_i[k_\mathrm{u}-1]\big]$ based on
  \begin{align}
 [\mathbf{y}_0\ \mathbf{y}_1\ \ldots\ \mathbf{y}_{i+T_\mathrm{u}}]=  [g_n(\mathbf{x}_0, e_0)\ g_n(\mathbf{x}_1, e_1)\ \ldots\ g_n(\mathbf{x}_{i+T_\mathrm{u}}, e_{i+T_\mathrm{u}})],\label{appendixBeq6}
 \end{align}
According to~\eqref{defXiAppendixB}, for each $i\in\{-n+1, -n+2, \ldots\}$, the symbols in $\big[x_i[0]\ x_{i+1}[1]\ \cdots \ x_{i+n-1}[n-1]\big]$ are transmitted between time~$i$ to $i+n-1$. Therefore, it follows from~\eqref{appendixBeq1}, Property~(iii) and~\eqref{appendixBeq5} that for each $i\in\mathbb{Z}_+$ and each $\ell\in\{0, 1, \ldots, k_\mathrm{v}-1\}$, the destination can perfectly recover $v_i[\ell]$ by time $i+T_\mathrm{v}$ based on $ [\mathbf{y}_0\ \mathbf{y}_{1}\ \ldots\ \mathbf{y}_{i+T_\mathrm{v}}]$. Similarly, it follows from~\eqref{appendixBeq1}, Property~(iv) and~\eqref{appendixBeq6} that for each $i\in\mathbb{Z}_+$ and each $\ell\in\{0, 1, \ldots, k_\mathrm{u}-1\}$, the destination can perfectly recover $u_i[\ell]$ by time $i+T_\mathrm{u}$ based on $ [\mathbf{y}_0\ \mathbf{y}_{1}\ \ldots\ \mathbf{y}_{i+T_\mathrm{u}}]$.  Consequently, for any $i\in\mathbb{Z}_+$ and any $\mathbf{e}\in\Omega_B$, the destination can perfectly recover $v_i[\ell]$ by time $i+T_\mathrm{v}$ for each~$\ell\in\{0, 1, \ldots, k_\mathrm{v}-1\}$ and perfectly recover $u_i[\ell]$ by time $i+T_\mathrm{u}$ for each~$\ell\in\{0, 1, \ldots, k_\mathrm{u}-1\}$, which then implies by Definition~\ref{definitionAchievability} that the $(n, k_\mathrm{v}, k_\mathrm{u}, T_\mathrm{v}, T_\mathrm{u})_{\mathbb{F}}$-streaming code corrects any $B$-erasure sequence.

\begin{thebibliography}{10}
\providecommand{\url}[1]{#1}
\csname url@samestyle\endcsname
\providecommand{\newblock}{\relax}
\providecommand{\bibinfo}[2]{#2}
\providecommand{\BIBentrySTDinterwordspacing}{\spaceskip=0pt\relax}
\providecommand{\BIBentryALTinterwordstretchfactor}{4}
\providecommand{\BIBentryALTinterwordspacing}{\spaceskip=\fontdimen2\font plus
\BIBentryALTinterwordstretchfactor\fontdimen3\font minus
  \fontdimen4\font\relax}
\providecommand{\BIBforeignlanguage}[2]{{%
\expandafter\ifx\csname l@#1\endcsname\relax
\typeout{** WARNING: IEEEtran.bst: No hyphenation pattern has been}%
\typeout{** loaded for the language `#1'. Using the pattern for}%
\typeout{** the default language instead.}%
\else
\language=\csname l@#1\endcsname
\fi
#2}}
\providecommand{\BIBdecl}{\relax}
\BIBdecl

\bibitem{CiscoWhitePaper2018}
Cisco, ``Cisco visual network index: Forecast and methodology, 2017-2022,''
  Tech. Rep., Nov. 2018.

\bibitem{tactileInternet16}
M.~Simsek, A.~Aijaz, M.~Dohler, J.~Sachs, and G.~Fettweis, ``{5G}-enabled
  {Tactile Internet},'' \emph{{IEEE} J. Sel. Areas Commun.}, vol.~34, no.~3,
  pp. 460--473, 2016.

\bibitem{5GPPP}
\BIBentryALTinterwordspacing
5G-PPP, ``{5G} empowering vertical industries,'' Tech. Rep., Feb. 2015.
  [Online]. Available: \url{https://5g-ppp.eu/roadmaps/}
\BIBentrySTDinterwordspacing

\bibitem{onewayTransTime}
{International Telecommunication Union}, ``Recommendation {G.114},'' Tech.
  Rep., May 2003.

\bibitem{StockhammerHannuksela2005}
T.~Stockhammer and M.~Hannuksela, ``{H.264/AVC} video for wireless
  transmission,'' \emph{{IEEE} Wireless Commun. Mag.}, vol.~12, pp. 6--13, Aug.
  2005.

\bibitem{BKTAmagazine17}
A.~Badr, A.~Khisti, W.-T. Tan, and J.~Apostolopoulos, ``Perfecting protection
  for interactive multimedia: {A} survey of forward error correction for
  low-delay interactive applications,'' \emph{IEEE Signal Processing Magazine},
  vol.~34, pp. 95 -- 113, 2017.

\bibitem{Bolot1993}
J.~Bolot, ``Characterizing end-to-end packet delay and loss in the
  {Internet},'' \emph{Journal of High Speed Networks}, vol.~2, no.~3, pp.
  305--323, 1993.

\bibitem{Paxson1999}
V.~Paxson, ``End-to-end {Internet} packet dynamics,'' \emph{{IEEE/ACM} Trans.
  Netw.}, vol.~7, no.~3, pp. 277--292, 1999.

\bibitem{Gilbert1960}
E.~N. Gilbert, ``Capacity of a burst-noise channel,'' \emph{Bell System
  Technical Journal}, vol.~39, pp. 1253--–1265, Sep. 1960.

\bibitem{Elliott1963}
E.~O. Elliott, ``Estimates of error rates for codes on burst-noise channels,''
  \emph{Bell System Technical Journal}, vol.~42, pp. 1977--1997, Sep. 1963.

\bibitem{Fritchman1967}
B.~D. Fritchman, ``A binary channel characterization using partitioned {Markov}
  chains,'' \emph{{IEEE} Trans. Inf. Theory}, vol.~13, no.~2, pp. 221--227,
  1967.

\bibitem{MartinianSundberg2004}
E.~Martinian and C.-E.~W. Sundberg, ``Burst erasure correction codes with low
  decoding delay,'' \emph{{IEEE} Trans. Inf. Theory}, vol.~50, no.~10, pp. 2494
  -- 2502, 2004.

\bibitem{LeungHo2012}
D.~Leong and T.~Ho, ``Erasure coding for real-time streaming,'' in \emph{Proc.
  IEEE Intl. Symp. Inf.~Theory}, Cambridge, MA, Jul. 2012.

\bibitem{LQH2013}
D.~Leong, A.~Qureshi, and T.~Ho, ``On coding for real-time streaming under
  packet erasures,'' in \emph{Proc. IEEE Intl. Symp. Inf.~Theory}, Istanbul,
  Turkey, Jul. 2013.

\bibitem{BKTA2013}
A.~Badr, A.~Khisti, W.-T. Tan, and J.~Apostolopoulos, ``Streaming codes for
  channels with burst and isolated erasures,'' in \emph{IEEE INFOCOM}, Turin,
  Italy, Apr. 2013.

\bibitem{AdlerCassuto2017}
N.~Adler and Y.~Cassuto, ``Burst-erasure correcting codes with optimal average
  delay,'' \emph{{IEEE} Trans. Inf. Theory}, vol.~63, no.~5, pp. 2848--2865,
  2017.

\bibitem{FKLTZA2018}
S.~L. Fong, A.~Khisti, B.~Li, W.-T. Tan, X.~Zhu, and J.~Apostolopoulos,
  ``Optimal streaming codes for channels with burst and arbitrary erasures,''
  \emph{{IEEE} Trans. Inf. Theory}, vol.~65, no.~7, pp. 4274 -- 4292, 2019.

\bibitem{QUIC2017}
{A.~Langley \textit{et al.}}, ``The {QUIC} transport protocol: {Design} and
  internet-scale deployment,'' in \emph{Proc.\ ACM SIGCOMM}, Los Angeles, CA,
  USA, Aug. 2017.

\bibitem{BLKTZA2018}
A.~Badr, D.~Lui, A.~Khisti, W.-T. Tan, X.~Zhu, and J.~Apostolopoulos,
  ``Multiplexed coding for multiple streams with different decoding delays,''
  \emph{{IEEE} Trans. Inf. Theory}, vol.~64, no.~6, pp. 4365 -- 4378, 2018.

\bibitem{MacWilliamsSloane1988}
F.~J. MacWilliams and N.~J.~A. Sloane, \emph{The Theory of Error-Correcting
  Codes}, 1st~ed.\hskip 1em plus 0.5em minus 0.4em\relax Amsterdam, Holland:
  North-Holland, Netherlands, 1988.

\bibitem{Forney1971}
G.~D. Forney, ``Burst-correcting codes for the classic bursty channel,''
  \emph{{IEEE} Trans. Inf. Theory}, vol.~19, no.~5, pp. 772 -- 781, 1971.

\bibitem{Cover06}
T.~M. Cover and J.~Thomas, \emph{Elements of Information Theory}, 2nd~ed.\hskip
  1em plus 0.5em minus 0.4em\relax Hoboken, NY: John Wiley and Sons Inc., 2006.

\bibitem{EKT2014}
F.~Etezadi, A.~Khisti, and M.~Trott, ``Zero-delay sequential transmission of
  {Markov} sources over burst erasure channels,'' \emph{{IEEE} Trans. Inf.
  Theory}, vol.~60, no.~8, pp. 4584–--4613, 2014.

\bibitem{BPKTA17}
A.~Badr, P.~Patil, A.~Khisti, W.-T. Tan, and J.~Apostolopoulos, ``Layered
  constructions for low-delay streaming codes,'' \emph{{IEEE} Trans. Inf.
  Theory}, vol.~63, no.~1, pp. 111 -- 141, 2017.

\end{thebibliography}
\end{document}